\documentclass[usenatbib]{mn2e}
\usepackage{epsfig}
\catcode`\@=11
\def\gta{\ifmmode{\,\mathrel{\mathpalette\@versim>\,}}
    \else{$\,\mathrel{\mathpalette\@versim>}\,$}\fi}
\def\lta{\ifmmode{\,\mathrel{\mathpalette\@versim<\,}}
    \else{$\,\mathrel{\mathpalette\@versim<}\,$}\fi}
\def\@versim#1#2{\lower 2.9truept \vbox{\baselineskip 0pt \lineskip
    0.5truept \ialign{$\m@th#1\hfil##\hfil$\crcr#2\crcr\sim\crcr}}}
\catcode`\@=12  
{\newif\ifnotend
\notendtrue
\def\veclist{ABCDEFGHIJKLMNOPQRSTUVWXYZabcdefghijklmnopqrstuvwxyz.}
\def\top#1#2.{#1}
\def\tail#1#2.{#2.}
\loop\expandafter\xdef\csname v\expandafter\top\veclist\endcsname%
{{\noexpand\bf\expandafter\top\veclist}}
\edef\veclist{\expandafter\tail\veclist}
\if\veclist.\notendfalse\fi\ifnotend\repeat}

\def\kms{\,{\rm km}\,{\rm s}^{-1}}
\def\mag{\,{\rm mag}}
\def\mas{\,{\rm mas}}

\def\Myr{\,{\rm Myr}}
\def\yr{\,{\rm yr}}
\def\Gyr{\,{\rm Gyr}}
 
\def\pc{\,{\rm pc}}
\def\kpc{\,{\rm kpc}}

\def\meh{\hbox{[Me/H]}}
\def\feh{\hbox{[Fe/H]}}
\def\llg{\log({\rm g})}

\def\dex{\,{\rm dex}}
\def\figref#1{Fig.~\ref{#1}}

\def\ovcv0{\overline{V_0}}

\newcommand {\Rsun}{{R_{0}}}
\newcommand {\vphisun}{V_{g, \odot}}
\newcommand {\Vtot}{V_{g, \odot}}

\newcommand {\Usun}{{U_{\!\odot}}}
\newcommand {\Usunzero}{{U_{{\!\odot},0}}}
\newcommand {\Wsun}{{W_{\!\odot}}}
\newcommand {\Vsun}{{V_{\!\odot}}}
 
\newcommand{\beq}{\begin{equation}}
\newcommand{\eeq}{\end{equation}}

\title[Galactic Rotation and the Solar Motion]{Galactic Rotation and Solar Motion from Stellar Kinematics}

\author[Ralph Sch\"onrich]{Ralph Sch\"onrich$^{1,2}$\thanks{E-mail:
    schonrich.1@osu.edu}\\
 $^{1}$ Max-Planck-Institut f\"ur Astrophysik, Garching, Germany \\ 
 $^{2}$ Hubble Fellow, Ohio State University, Columbus, OH 43210 \\
}

\begin{document}

\date{accepted July 2nd, 2012}

\pagerange{\pageref{firstpage}--\pageref{lastpage}} \pubyear{2012}

\maketitle

\label{firstpage}

\begin{abstract}
I present three methods to determine the distance to the Galactic centre $\Rsun$, the solar azimuthal velocity in the Galactic rest frame $\Vtot$ and hence the local circular speed $V_c$ at $\Rsun$. These simple, model-independent strategies reduce the set of assumptions to near axisymmetry of the disc and are designed for kinematically hot stars, which are less affected by spiral arms and other effects. 

The first two methods use the position-dependent rotational streaming in the heliocentric radial velocities ($U$). The resulting rotation estimate $\theta$ from $U$ velocities does not depend on $\Vtot$. The first approach compares this with rotation from the galactic azimuthal velocities to constrain $\Vtot$ at an assumed $\Rsun$. Both $\Vtot$ and $\Rsun$ can be determined using the proper motion of Sgr $A^{*}$ as a second constraint. The second strategy makes use of $\theta$ being roughly proportional to $\Rsun$. Therefore a wrong $\Rsun$ can be detected by an unphysical trend of $\Vtot$ with the intrinsic rotation of different populations. From these two strategies I estimate $\Rsun = (8.27 \pm 0.29) \kpc$ and $\Vtot = (250 \pm 9) \kms$ for a stellar sample from SEGUE, or respectively $V_c = (238 \pm 9) \kms$. The result is consistent with the third estimator, where I use the angle of the mean motion of stars, which should follow the geometry of the Galactic disc. This method also gives the Solar radial motion with high accuracy.

The rotation effect on $U$ velocities must not be neglected when measuring the Solar radial velocity $\Usun$. It biases $\Usun$ in any extended sample that is lop-sided in position angle $\alpha$ by of order $10 \kms$. Combining different methods I find $\Usun \sim 14 \kms$, moderately higher than previous results from the Geneva-Copenhagen Survey.

\end{abstract}

\newcounter{mytempeqncnt}

\section{Introduction}

Among the central questions in Galactic structure and parameters are the Solar motion, Solar Galactocentric radius $R_0$ and the local circular velocity $V_c$ of our Galaxy. Galactic rotation curves are found to be generally quite flat over a vast range of Galactocentric radii \citep[][]{Krumm79}. As common for Galaxies with exponential discs \citep[][]{Freeman70} there is some evidence for a radial trend of the Galactic circular velocity near the Sun, but it is very moderate \citep[][]{Feast97, McMillanB09}, so that the circular velocity at solar Galactocentric radius $\Rsun$ characterises well the entire potential. 

Initially local kinematic data from stellar samples were the main source to extract Galactic parameters including $V_c$, e.g. by the use of the Oort constants \citep[][]{Oort27}. The kinematic heat of stellar populations requires large sample sizes, so that studies determining the Local Standard of Rest (LSR) are still primarily based on stars, but some classical strategies like the position determination of the Galactic centre pioneered by \cite{Shapley18} reached their limits by the constraints from geometric parallaxes, by the number of available luminous standard candles, and by the magnitude requirements of stellar spectroscopy. 

Apart from some more recent attempts to use luminous stars \citep[e.g.][]{Burton74}, most current evidence on $\Rsun$ and $V_c$ derives from modelling streams in the Galactic halo \citep[see e.g.][]{Ibata01, Majewski06} and from radio observations of the $HI$ terminal velocity \citep[see][]{McMillan11}, the Galactic centre, molecular clouds and MASERs \citep[e.g.][]{Reid04}. While the first branch relies strongly on assumptions such as distance scale and shape of the Galactic potential \citep[cf. the discussion in ][]{Majewski06}, there are further complications like the failure of tidal streams to delineate stellar orbits \citep[][]{EyreB09}. 

Radio observations \citep[][]{Reid04} have accurately determined the proper motion of the radio source Sgr $A^*$, which is identified with the central black hole of the Milky Way \citep[for a discussion of possible uncertainties see also][]{Broderick11}. It tightly constrains the ratio of the solar speed in the azimuthal direction $\Vtot$ to $\Rsun$, but further information is required to obtain both quantities, letting aside the need for an independent measurement. Recently parallaxes to objects in the central Galactic regions have become available \citep[see the discussion in][]{Reid09}, and values for the Galactic circular speed have been derived from $HI$ motions and from MASERs \citep[][]{Rygl10}. Despite the decent errors in the determined kinematics of MASERs, the small sample sizes impose considerable systematic uncertainties: they are not on pure circular orbits and more importantly they are intimately connected to the intense star formation in spiral arms, where the kinematic distortions are largest. As the youngest strategy, studies of orbits in the Galactic centre have gained high precision, but weakly constrain $R_0$ due to a strong degeneracy with the black hole mass \citep[][]{Ghez09}. Hence the values for $\Rsun$ and $\Vtot$ remain under debate. 

An independent determination of Galactic parameters is facilitated by the new large spectroscopic surveys like RAVE \citep[][]{RAVEI} and SEGUE \citep[][]{Yanny09}. I will show that already now the stellar samples, which so far have been primarily used for the exploration of substructure \citep[][]{Belokurov07, Hahn11} give results for $V_c$ competitive with radio observations.

Common ways to determine Galactic parameters from the motion of stars require a significant bundle of assumptions and modelling efforts to evaluate the asymmetric drift in a subpopulation compared to the velocity dispersion and other measurements and assumptions, e.g. detailed angular momentum and energy distributions, the validity of the theoretical approximations, etc. On the contrary a good measurement strategy for Galactic parameters should have the following properties:
\begin{itemize}
\item Do not rely on dispersions and other quantities that require accurate knowledge of measurement errors.
\item Do not rely on kinematically cold objects in the disc plane, which are particularly prone to perturbations from Galactic structure.
\item Do not rely on any specific models with hidden assumptions and parameters and require as few assumptions as possible. 
\end{itemize}

In an attempt to approximate these conditions this note concentrates on the use of kinematically hot (thick disc, halo) populations. The use of mean motions and the assumption of approximate axisymmetry in our Galaxy will be sufficient to fix the Galactocentric radius of the Sun as well as the Solar motion and Galactic circular speed.

In Section \ref{sec:p7general} I lay out the general method before describing the sample selection and treatment in Section \ref{sec:p7sample}. The latter comprises a discussion of proper motion systematics, a discussion of distances in subsection \ref{sec:p7dist} and a discussion of line-of-sight velocities and the vertical motion of the Sun. In Section \ref{sec:p7globrot} I lay out the radial velocity based rotation measurement using SEGUE, discuss the radial velocity of the Sun in Subsection \ref{sec:p7LSR} and shortly discuss possible trends. From \ref{sec:p7Galparam} on I present three methods to to extract Galactic parameters from stellar samples, the first two methods relying on the radial velocity based rotation measurement and the third on the mean direction of stellar motions. In Section \ref{sec:p7conclude} I summarise the results.

\section{General Idea}\label{sec:p7general}
Before I start the dirty work of data analysis it seems appropriate to concisely lay out the general definitions and ideas.

\begin{figure}
\epsfig{file=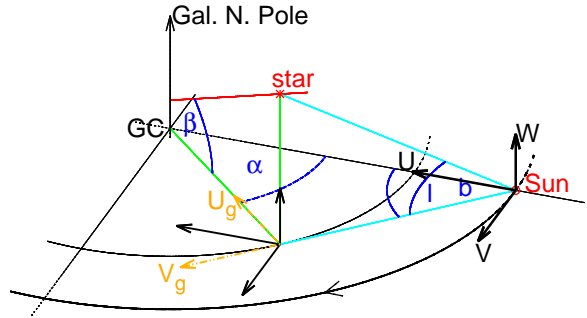,angle=-90,width=\hsize}
\caption[Definition of kinematic quantities]{Definition of kinematic and geometric quantities. On the connection line between Sun and Galactic centre (abbreviated GC) the heliocentric ($U,V,W$) and galactocentric velocities ($U_g,V_g,W_g$), i.e. velocities in the galactocentric cylindrical coordinate system are identical, while in general they differ. This gives rise to systematic streaming motion in the heliocentric frame.
}\label{fig:p7geo}
\end{figure}

\subsection{Definitions}\label{sec:p7def}

This work is based on comparisons between velocities in the heliocentric frame and galactocentric velocities. In the heliocentric frame I define the velocities $U, V, W$ in a Cartesian coordinate system as the components pointing at the Sun towards the Galactic centre, in the direction of rotation and vertically towards the Galactic North pole. The galactocentric velocities are defined in cylindrical coordinates around the Galactic centre with $U_g, V_g, W_g$ pointing again towards the Galactic central axis, in the direction of rotation and vertically out of the plane, as illustrated in \figref{fig:p7geo}. $\Usun, \Vsun, \Wsun$ are the three components of the Solar motion relative to the Local Standard of Rest (i.e. the circular orbit at the local galactocentric radius $R_0$), for which I assume the values from \cite{SBD} if not stated otherwise. The total azimuthal velocity of the Sun in the Galactic frame is written $\vphisun = \Vsun + V_c$ where $V_c$ denotes the circular velocity in the disc plane of the Milky Way at $R_0$. The mean rotation speed of a subsample of stars will be named $\theta$ and by the asymmetric drift is generally smaller than $V_c$. For convenience local Cartesian coordinates are defined left-handed $x,y,z$ pointing outwards, in the direction of rotation and upwards with the Sun at the origin.

\subsection{An absolute measure of rotation}\label{sec:p7gendisc}

Commonly studies on the rotation of Galactic components suffer from the uncertainty in the solar azimuthal velocity that directly translates into a systematic uncertainty of stellar azimuthal velocities. There is a way out: As already discussed in \cite{SBA} and evident from \figref{fig:p7geo}, Galactic rotation leaves its imprint in both the heliocentric $U$ and $V$ velocities. The direction of the rotational component of motion $V_g$ turns with the angle $\alpha$ between the lines Galactic centre -- Sun and Galactic centre -- star to align partly with the radial component $U$ of the local Cartesian frame. From this systematic streaming motion in heliocentric $U$ velocities we can directly infer the Galactic rotation of a stellar sample. 
Accounting for rotation the mean motions of stars in the heliocentric frame are:
\begin{eqnarray}\label{eq:p7rot}
\langle U \rangle &=& -\Usun + \theta \sin\alpha \nonumber \\ 
\langle V \rangle &=& -\Vtot + \theta(1 - \cos\alpha) \\ 
\langle W \rangle &=& -\Wsun ,\nonumber
\end{eqnarray}
with the rotation speed of the population $\theta$ and the Galactic angle $\alpha$. $(\Usun,\Vtot,\Wsun)$ are the velocity components of the Solar motion in the Galactic rest frame radially towards the Galactic centre, azimuthally in the direction of disc rotation and vertically out of the plane. Observing the change of heliocentric $U$ velocity in a sufficiently extended sample hence estimates the rotation of a component $\theta$, once we know the angle $\alpha$.

Similarly the observed heliocentric azimuthal velocity $V$ shrinks towards larger $|\sin \alpha|$. Prima facie it might be tempting to use this term. Yet, all it expresses is a slowing of heliocentric azimuthal velocities. In general $\theta$ is not a real constant, but a function of altitude $z$, galactocentric radius $R$, metallicities, etc. In particular the kinematically hotter disc populations at higher altitudes above the plane have a larger asymmetric drift \citep[see e.g.][]{Binney10}, or vice versa slower rotation, seconded by an increase in the share of the nearly non-rotating halo. Pointing out from the Sun's position in the mid-plane, $|\cos\alpha|$ correlates via the typical distance range with altitude $z$. In contrast to $U$, where this bias is of second order, it is of first order in $V$ and prohibits any naive estimate of $\theta$ from azimuthal velocities. A similar uncertainty derives from the possibility for variations of $\theta$ in $R$ even at the same altitude. 

The radial velocities are better behaved. Sample changes along the baseline $\sin\alpha$ are not only a second order effect, but are generally more robust: the greatest concern is a bias in the derived $\Usun$, which can be prevented by fixing its value by previous knowledge or by simply having a sufficient sky coverage that allows to balance both sides of the Galactic centre and includes sufficient numbers of stars at low $|\sin \alpha|$. If the integrity of $\Usun$ is ensured, sample changes along the baseline just result in different statistical weights of the involved populations.

\begin{figure}
\epsfig{file=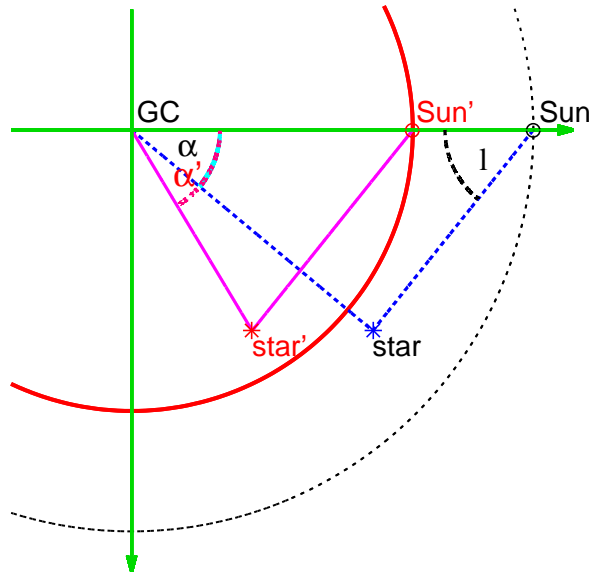,angle=-90,width=\hsize}
\caption[Varying the solar position]{Changing the Galactocentric radius of the Sun $R_0$. The Galactic longitude of the star does not change so that the angle $\alpha$ grows with shrinking $R_0$. This reduces the rotation speed estimated from heliocentric radial velocities proportionally.
}\label{fig:p7varr}
\end{figure}

The value $\theta$ derived from heliocentric radial motions is by construction independent from the assumed Solar azimuthal velocity and (in an extended sample) is only weakly impacted by (if held fixed) a mis-estimate of $\Usun$. This contrasts to direct measurements of $V_g$ that rely fully on the total solar azimuthal velocity $\Vtot$ - a short-come now turning into a virtue: $\Vtot$ and $\Rsun$ can be varied to force an agreement between the rotation estimates from $\theta$ and $V_g$. As the position-dependent statistical weights in $\theta$ are known, one can estimate a weighted average of $V_g$ exactly on these weights. The sole assumption of approximate axisymmetry of the Galactic disc (but not necessarily in the sample selection) is then sufficient to compare $\theta$ and $V_g$ and measure Galactic parameters without the least concern about sample selection, velocity trends, etc. The comparison gives a relation between $\Vtot$ and $\Rsun$, which can be combined with independent data like the proper motion of Sgr $A^*$ to provide a superior estimate of Galactic parameters. However, $\Rsun$ alone can be measured with this simple method alone, as $\theta$ is roughly proportional to $\Rsun$ (see Section \ref{sec:p7Galparam} or \figref{fig:p7varr}), so that any error in $\Rsun$ shears the estimate of $\Vtot$ for fast rotating populations against their slower counterparts.

\section{Sample selection and distances}\label{sec:p7sample}

\subsection{Data selection}

Any reader not interested into the details of sample selection, distance analysis and the proper motions may skip this section and turn directly towards Section \ref{sec:p7LSR}.

All data used in this study were taken from the seventh and eighth data release \citep[hereafter DR7 and DR8][]{SloanDR7, SloanDR8} of the Sloan Digital Sky Survey \citep[SDSS][]{SDSS3}. The stellar spectra stem from the Sloan Extension for Galactic Understanding and Exploration \citep[SEGUE,][]{Yanny09}. An evaluation of the performance of their parameter pipeline (SSPP) can be found in \cite{Lee08a, Lee08b}. 
This work has no interest in specific metallicity distributions, but in having kinematically unbiased samples that include as many stars as possible. Hence the sample drawn from DR8 consists of a raw dataset of $224\,019$ stars from all SEGUE target categories and the photometric and reddening standard stars that do not include any proper motion cuts and have clean photometry. In particular I use the categories of F turnoff/sub-dwarf, Low Metallicity, F/G, G dwarf, K dwarf, M sub-dwarf stars from SEGUE1, as well as MS turnoff, Low Metallicity and the reddening and photometric standard stars from all samples. The categories were selected by the target bitmasks of the database. Some of these categories overlap with kinematically biased selections and the survey descriptions are not clear about a possibility of direct (cross-selection of stars) or indirect (non-selection of stars that are targeted in a biased category) bias.\footnote{cf. the comments of \cite{Schlesinger11} at the analogous problem of the metallicity distribution.} I made sure that the presented results are unchanged when selecting stars by the unique target flags in the SSPP. Also checks on leaving out e.g. K dwarfs that might overlap with the biased K giants did not provide evidence for a notable bias by the SEGUE target selection. 
The older DR7 forms a subset of $182627$ stars. 

To ensure sufficient quality of the involved proper motions, I select only stars with a match in the proper motion identifications ($match = 1$), the presence of at least $5$ detections ($nFit \ge 5$) and a good position determination with $\sigma_{RA}, \sigma_{DEC} < 500 \,{\rm mas}$ as a compromise between \cite{Munn04} and currently used selection schemes. I further require the formal errors on the proper motions to fulfil $\sigma_{\mu_b}, \sigma_{\mu_l} < 4 \,{\rm mas\,yr}^{-1}$, removing the tail of uncertain proper motion determinations and require line-of-sight velocity errors below $20 \kms$.

Sample homogeneity is not a concern in this study, but to avoid issues with reliability of the isochrones in colour regions with insufficient coverage to check them by statistics, I select $0 < (g-i)_0 < 1$, which tosses out a handful of stars especially on the red side. I tested that narrower colour cuts would have no significant effect on the presented results.

For our purpose it is most important to have a good discrimination between dwarf stars and other categories and to dispose of as many evolved stars above the main sequence as possible. Hence only stars are used that have values for gravity and metallicity determined by the pipeline. The sample must be restricted to dwarf stars, as \cite{SBA} showed that stars with intermediate gravities in the pipeline are an indeterminable blend of dwarf, subgiant and giant stars, which renders their distance scatter and residual biases detrimental to any investigation of their kinematics.

Following \cite{SBA} I adopt a sloping gravity cut generally tighter than the usually applied $\llg > 4.0$, demanding
$\llg > 4.2 + 0.15\feh$ and setting a constant limit of $\llg > 3.9$ for $\feh < -2$ and $\llg > 4.2$ for $\feh > 0$. It is important to point out that this selection could especially on the metal-poor end still be irresponsibly lenient for studies of detailed stellar kinematics, as it will comprise a serious contamination with misidentified (sub-)giants and does not bias sufficiently against the brighter turn-off stars. Indeed the distance corrections described below react to a tighter gravity cut in the metal-poor regime by up to $10 \%$. Relying only in mean velocities the preference in this work is leaned towards larger sample numbers and for assurance I checked that a tighter selection widens the error margins, but does not alter my results.

Quite frequently studies on SEGUE use a cut in the signal to noise ratio $S/N > 10$ to ensure reasonable parameters. Since the determination of stellar parameters is difficult in those noisy spectra, the detection of a mild drift in the derived quantities towards higher $S/N$ does not surprise, motivating a cut at $S/N > 15$ where results become more stable.

De-reddening is done by subtraction of the reddening vectors given in the DR8 catalogue from the observed magnitudes. To limit uncertainties and avoid problems with complicated three dimensional behaviour of the reddening, I remove all objects with an estimated $g$ magnitude extinction $A_g > 0.75 \mag$. This strips all stars in the low latitude fields from the sample. I tested that the exclusion does not have a significant impact on my results apart from mildly increasing the derived formal errors, but maintain this cut for the sake of sample purity. I exclude all stars flagged for spectral anomalies apart from suspected carbon enhancement. The internal consistency of the distance estimator against closer distance bins starts eroding for $s > 3.6 \kpc$. Hence I use this distance limit instead of the more common $s < 4.0 \kpc$.

\subsection{Proper motions}\label{sec:p7pm}

For a precise measurement of Galactic rotation I am reliant on a decent control of systematic errors on proper motions. Unbiased noise is of no concern as proper motion errors translate linearly into velocities \citep[see equ. 20 in][]{SBA} and hence mean velocities (on which this paper relies) are only affected by the mean value of proper motion errors.\footnote{The knowledge of proper motion errors themselves plays only a subordinate role in the distance correction} After a short description of proper motion systematics I will measure them in equatorial coordinates in Section \ref{sec:p7pmeq} and develop terms that are more natural to a contamination problem and Galactic rotation in Section \ref{sec:p7pmgal}.

\cite{Sloanerr} reported significant problems with the DR8 astrometry implying also problems with proper motions. Fortunately the revised astrometry has become available during refereeing stage of this paper and I will exclusively use the new data. It should be noted that also DR7 had major problems, e.g. the entire $rerun$ $648$ was clearly erroneous/contaminated with large effects on kinematics. Its $15\%$ share of the total sample would bring down my later measurement of Galactic rotation by as much as $14 \kms$. Most problems with Sloan proper motions have been adressed in the revised astrometry, but the large impact calls for an investigation of residual biases. 

Two main sources of systematic terms can be identified: astrometric ``frame-dragging'' and chromatic aberrations. Despite the efforts to clean up the Galaxy sample used for astrometry from residual contamination with Galactic stars, the sample may still be subject to some astrometric ``frame-dragging''. Qualitatively those stars shift the frames in the direction of Galactic rotation or respectively their motion relative to the Sun, underestimating the real motion in the sample. As an alternative explanation \cite[][]{Bond09} traced the significant declination dependent net proper motions on quasars back to chromatic aberration by Earth's atmosphere \citep[][]{Kaczmarczik09}: the angle at which the stellar light passes through our atmosphere strongly depends on the declination and this should give rise to colour-dependent offsets. Further, deformations in the telescope can introduce minor distortions of the focal plane. 

\begin{table}
\begin{math}
\begin{array}{l|ccc}
 & $ fitting $ \mu_{\rm RA} \quad
& $ fitting $ \mu_{\rm DEC}\quad & $     error    $ \\\hline
f_{i,{\rm RA}} \cdot 10^3 &  ${\bf 0.33}$ & 0.18 & 0.16\parbox[0pt][1.5em][c]{0cm}{}\\
f_{i,{\rm DEC}}\cdot 10^3 & -0.59 & ${\bf 7.88}$ & 0.50\parbox[0pt][1.5em][c]{0cm}{} \\
c_i & ${\bf 0.115}$  & ${\bf -0.205}$ & 0.033 \parbox[0pt][1.5em][c]{0cm}{}
\end{array}
\end{math}
\caption{Parameters for systematic proper motion terms when fitting the linear equation (\ref{eq:p7RADEC}) to detect systematic proper motions on the quasar sample. Both declination (right column) and right ascension (left column) show significant systematics. Terms with $> 2 \sigma$ significance are printed in bold letters.}\label{tab:qparRADEC}
\end{table}

\subsubsection{Measuring proper motion systematics in equatorial coordinates}\label{sec:p7pmeq} 

While the existence of these aberrations appears plausible, I tested proper motions on the cleaner \cite{Schneider10} quasar sample. I fitted a simple linear function in right ascension and declination to the proper motions from the revised DR8 astronomy:
\begin{equation}
\frac{\mu_i}{{\rm mas}\,{\rm yr}^{-1}} = f_{i, \rm RA}\cdot \frac{\rm RA}{deg} + f_{i,\rm DEC} \cdot \frac{{\rm DEC}}{deg} + c_i 
\end{equation}\label{eq:p7RADEC}
where $i = {\rm RA},{\rm DEC}$ denotes the two possible proper motion components in right ascension (RA) and declination (DEC), $f$ are the coefficients, $c$ are free fitting constants. Using the combined fit diminishes the impact from omitted variable bias that might arise from the particular sample geometry (the explaining variables are mildly correlated). The fit parameters are shown in Table \ref{tab:qparRADEC}. There is a clear and significant detection of systematic proper motions of order $0.2 {\rm mas}\,{\rm yr}^{-1}$. I also checked the same fit on the older DR8 astrometry, where the derived terms were even slightly smaller. The large term on $f_{\rm DEC, DEC}$ can be expected from chromatic aberration. However, the strong effect on the right ascension is more difficult to explain by atmospheric influences, but may still come from specific distortions on the telescope or if the observations were made in a very particular manner that correlates the airmass of a stellar observation significantly to right ascension. The sample size of just about $31500$ quasars limits the possibility to dissect into subgroups. However, cutting in observed colours there is no detectable difference between quasars at $(g-r) < 0.1$ and $(g-r) > 0.3$ implying that in the colour range interesting for the stellar sample the bias is quite consistent. Derived values go only adrift for $(g-r) < 0$, which bears no importance to the stellar sample.

\subsubsection{Proper motion systematics linked to rotation}\label{sec:p7pmgal}

The strong deviations on the right ascension motivate tests for astrometric ``frame'' dragging by stellar contamination. In the following I develop a very crude analytic approximation for a proper motion bias in the $(l,b)$-plane as it could arise from Galactic streaming - note also that this toy is not used in any other way throughout the paper apart from the proper motion corrections: candidates for contamination of the galaxy sample should preferentially populate a certain magnitude range and also have preferred colours resembling the galaxies used. It hence seems appropriate to assume for simplification some dominant distance for stars contaminating the sample, placed at $s = 1 \kpc$. The outcomes will anyway not be critically affected by this distance (which can alter the shape of the systematic bias once the stars are sufficiently remote for the small angle approximation to break down). On this shell I assume the stars to rotate dependent on their altitude $z$ around the Galactic centre with $\theta(z) = (215 - 10 z/\kpc) \kms$. The proper motion can be written down from the heliocentric velocities from equation (\ref{eq:p7rot}) subtracting the reflex motions of the solar Local Standard of Rest (LSR) values from \cite{SBD} and assuming a total solar azimuthal velocity according to the IAU recommendations at $v_{\phi,\odot} = 232 \kms$. Using this the simple Galactic streaming terms are:
\begin{eqnarray} \label{eq:p7givesmucorrect}
\mu_l &=& \left( f_l \chi_l  + a_l \vphisun \cos l \cos b \right)/(sk) + c_l {\rm mas}\,{\rm yr}^{-1} \\ \nonumber
\mu_b &=& \left( f_b \chi_b + a_b \vphisun \sin b \sin l \right)/(sk) + c_b {\rm mas}\,{\rm yr}^{-1}
\end{eqnarray}
with
\begin{eqnarray}
\chi_l &=& -U \sin l + V \cos l \\ \nonumber
\chi_b &=& \sin b (U \cos l - V \sin l) - \Wsun \cos b
\end{eqnarray}
where $k \sim 4.74 \kms/{(\pc \, {\rm mas}\,{\rm yr}^{-1})}$ does the unit conversion, $U$ and $V$ are the heliocentric velocities from streaming minus the assumed motion of the Sun and $f_l, a_l, c_l, f_b, a_b, c_b$ are the fit parameters. The terms connected to $a_i$ give a crude estimate for the effects from halo streaming.

\begin{table}
\begin{tabular}{l|cccc}
name & DR8 & $\sigma_{\rm DR8}$ & $DR8_{(g-r)>0}$ & $\sigma_{{\rm DR8}, (g-r)>0}$\\\hline
$f_l$ &  $-0.0585$ & $0.0051$ & $-0.0615$ & $0.0055$\\
$c_l$ & $-0.0005$ & $0.011$ & $0.001$ & $0.012$ \\
$a_l$ & $0.00415$  & $0.00053$ & $0.00434$ & $0.00057$\\
$f_b$ & $-0.0589$ & $0.0054$ & $-0.0604$ & $0.0057$\\
$c_b$ & $-0.159$& $0.013$ & $-0.159$ & $0.014$ \\
$a_b$ & $0.00370$ & $0.00067$ & $0.00418$ & $0.00072$\\ 
\end{tabular}
\caption{Fit parameters for systematic streaming in the quasar proper motions in DR8 with and without cut for $(g-r) > 0$ as described in the text.}\label{tab:qpar}
\end{table}




The derived values for DR8 are presented in Table \ref{tab:qpar}. The old, erroneous DR8 and DR7 produce similar results. Slicing the sample in colour I could not detect any significant changes, just the bluemost $10\%$ of the quasars appear to be different. Still, cutting for $(g-r) > 0$ does not evoke notable consequences $10\%$, as well as removing the $10 \%$ of objects with the largest proper motions does not have any impact. 

The coefficients $f_i$ and $a_i$ may be understood roughly as contamination fractions, but not on accurate terms because the distance dependence of proper motions effects a strong dependence of the fractions on the assumed distance of stellar contaminants. The sign in the expected trends is not predictable a priori, as the \cite{Schneider10} sample may harbour itself some residual stellar contamination that competes with possible astrometry problems. In both cases the main rotation term $f_i$ is significantly negative hinting to some more misclassified stars affecting astrometry as compared to the quasar sample. On the other hand the coefficients $a_i$ that resemble terms expected for the Galactic halo are slightly positive, which might be interpreted as a tiny surplus of halo stars hiding in the quasar sample. I rather understand it as an omitted variable bias: Neither do I know the exact functional shapes and distance distributions of the contamination, nor do I have precise terms at hand to cope with the possible chromatic aberrations. Exact terms for contamination would demand a complete modelling of the entire measurement process including a reasonable model to the Milky Way, other galaxies and the exact Sloan selection function. If I fit alone either $a_i$ or $f_i$, both variables turn out negative and still their counterparts are negative when I fit them with this naively derived value. So one could argue that $f_i$ is getting excessively negative being balanced by a slightly positive $a_i$ to cope with these uncertainties. 

In the end I cannot achieve more than a rather phenomenological correction without being able to determine the true cause for the observed trends. The naive approximation still fulfils its purpose to parametrise the influence on Galactic rotation measurements. Throughout the following paper I will use the terms for $(g-r)>0$ to correct stelar proper motions independent of colour, since no significant colour dependence was detected. 

For proper motion errors, which influence the distance determinations, I use the pipeline values and checked that using the values from \cite{Dong11} for DR7 instead does not significantly affect the presented analysis.

\subsection{Distances}\label{sec:p7dist}

In this work I make use of the statistical distance correction method developed by \cite{SBA}. The approach relies on the velocity correlations over Galactic angles that are induced by systematic distance errors and entails a better precision and robustness than classical methods. The method can find the average distance in any sample of a couple of hundred stars. The only thing required as first input is a smooth distance or respectively absolute magnitude calibration that has a roughly correct shape in parameters like colours and metallicity. The shape has some importance as erroneous assumptions would result in an increased distance scatter on the sample bins and thus in residual systematic trends for different subpopulations.
The use of statistical distance corrections is mandatory as there are in all cases expected offsets by reddening uncertainties, age uncertainties, the uncertain binary fraction, mild differences by evolutionary differences on the main sequence, systematic metallicity offsets and by the helium enhancement problem, where the isochrones appear to be too faint at a fixed colour.

\subsubsection{First guess distances}

Primary stellar distances are derived from a dense grid of BASTI isochrones \citep[][]{Piet04, Piet06} that was kindly computed by S. Cassisi for the purpose of our age determinations in \cite{CSA11}. I use the $11.6 \Gyr$ isochrones which seems an appropriate compromise between the suspected age of the first disc populations \citep[see e.g.][]{AB09, SBI, Bensby04} and the age of halo stars. For metal-poor stars with $\feh < -0.6$ I account for alpha enhancement by raising the effective $\feh$ by $0.2 \dex$ as suggested by \cite{Chieffi91} or \cite{Chaboyer92}. In analogy to the observations of \cite{Bensby07,Melendez08} I let the alpha enhancement go linearly to zero towards solar metallicity: 
\begin{eqnarray}
\meh = \left\lbrace \begin{array}{lcl}
	   \feh + 0.2 & $ for $ & \feh < -0.6 \\
	  \feh - \frac{1}{3}\feh & $ for $ & -0.6 \le \feh \le 0 \\
	  \feh & $ for $ & 0 < \feh  \end{array} \right.
\end{eqnarray}

Some of the stars are located bluewards of the turn-off point at this age. To cope with these objects, I search the point at which the old isochrone is $0.5 \mag$ brighter than the corresponding $50 \Myr$ isochrone and extrapolate bluewards parallel to the main sequence defined by the $50 \Myr$ isochrones. I also experimented with using the turnoff point, i.e. the blue-most point of the old isochrone, but this introduces larger shot noise in distances that would hamper the distance corrections via the method of \cite{SBA}. Comparing the expected $r$-band absolute magnitude of each star to the measured reddening corrected $r_0$, I then infer its expected distance modulus and hence distance $s_0$. 

\begin{figure}
\begin{center}
\epsfig{file=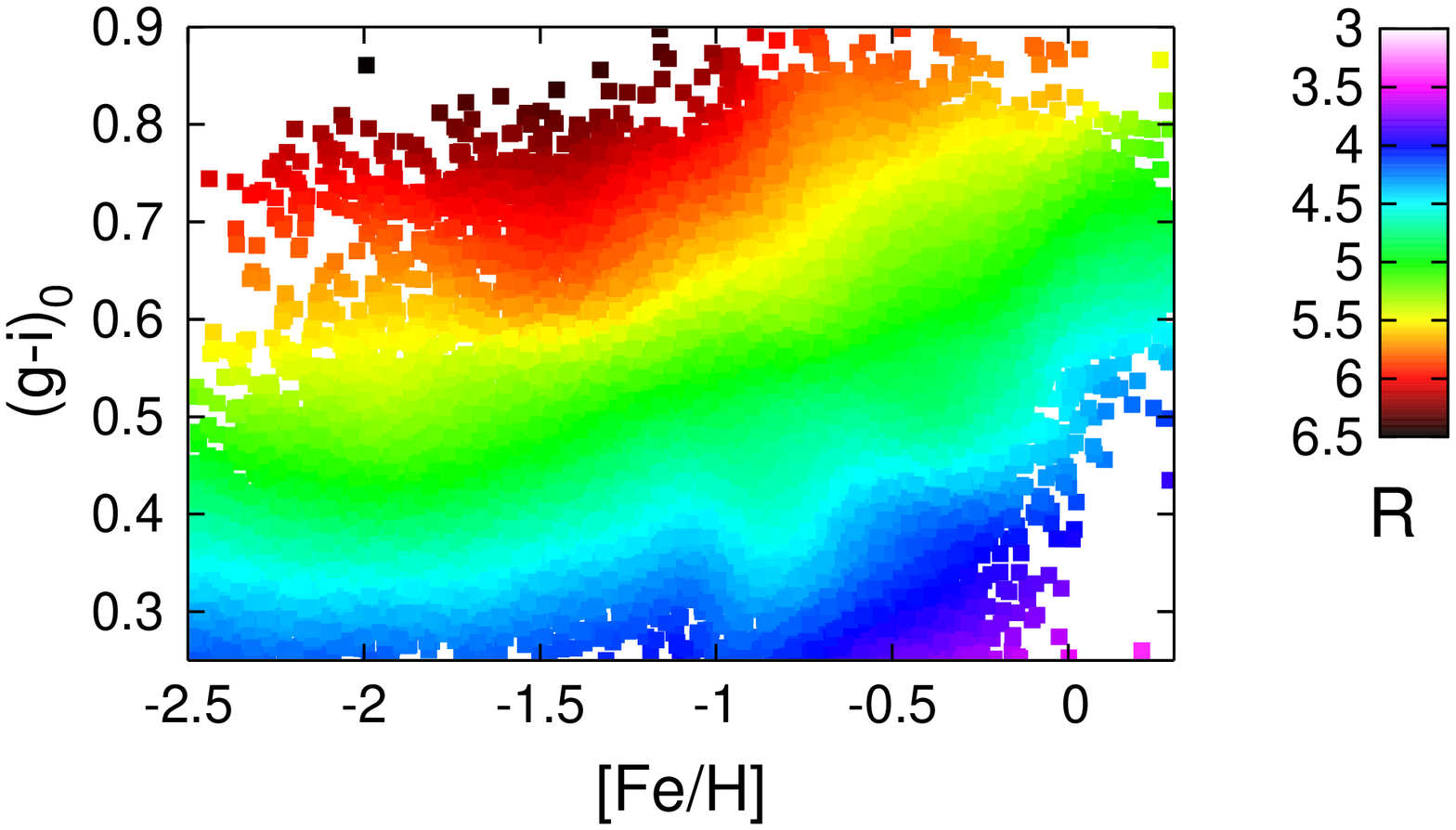,angle=-90,width=\hsize}
\epsfig{file=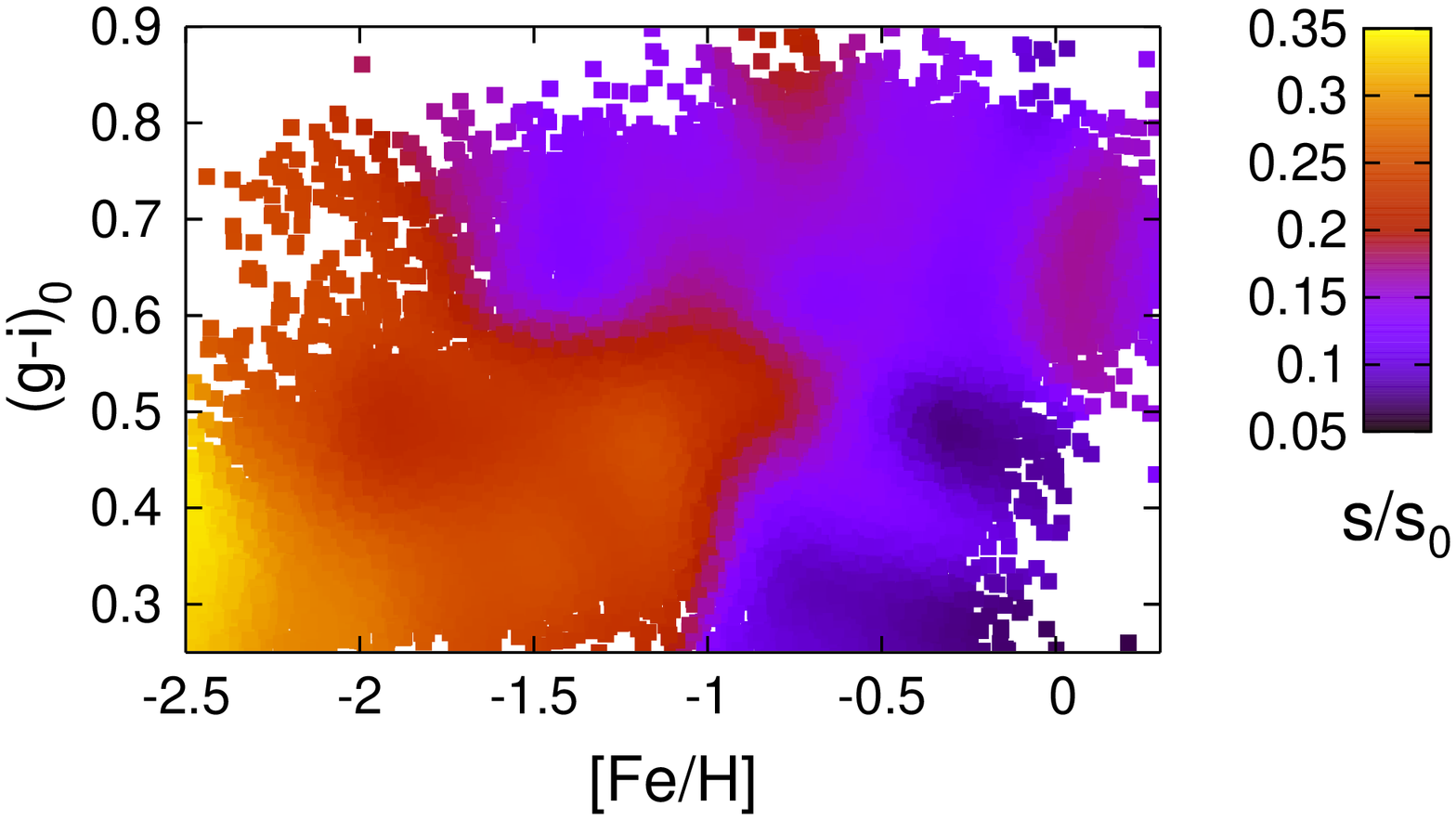,angle=-90,width=\hsize}
\end{center}
\caption[Assumed absolute $r$-band magnitudes from isochrone interpolation in the metallicity-colour plane after statistical distance correction and the distance correction factors]{Upper panel: Assumed absolute $r$-band magnitudes $R$ from isochrone interpolation in the metallicity-colour plane after statistical distance correction. Lower panel: Distance correction factors $s/s_0$ against the a priori distances $s_0$ in the metallicity-colour plane with the smoothing from equation (\ref{eq:p7gets}).
}\label{fig:p7distmap}
\end{figure}

\subsubsection{Distance corrections}

To derive a correction field, I bin the sample in metallicity and $(g-i)_0$ colour. I also attempted binning the sample simultaneously in distance as well, but found no additional benefits, probably because the bulk of the sample is further away than $0.5 \kpc$. I also excluded high reddening regions so that variations of reddening with distance are comparatively small.

There is a trade-off between resolution of the smallest possible structures in the correction field and the introduction of Poisson noise by lower effective numbers of stars contributing to the estimate at a certain point in the observed space. The strategy is to select every 30th star in the sample and estimate the distance shift for its 700 closest neighbours (I checked that the bin size has no significant effect) using an euclidean metric on the parameter space:
\begin{equation}
d_{\rm param}^2 = 2 \Delta_{{\rm(g-i)}_0}^2 + \Delta_{\feh}^2 $.$
\end{equation}
For non-halo or respectively metal-rich stars, which offer less accurate statistics by their smaller heliocentric velocities, this results in some significant scatter, which is reduced by smoothing the distance corrections via a weighted logarithmic mean:
\begin{equation}\label{eq:p7gets}
s = s_0 \exp{\left(\frac{\sum_{i}{\ln\left({1+x_{{\rm est}, i}}\right) \exp{\left(-\sum_j\frac{\Delta_{i,j}^2}{2\sigma_{j}^2}\right)}}}
{\sum_i{\exp{\left(-\sum_j\frac{\Delta_{i,j}^2}{2 \sigma_{j}^2}\right)}}}\right)} ,
\end{equation}
where $s_0$ is the a priori distance estimate from isochrone interpolation. The first sum runs over all stellar parameter sets $i$ around which the best-fit distance correction $x = 1 - 1/(1+f)$ or respectively the distance mis-estimate factor $f$ has been evaluated. The second index $j$ runs over used parameters (here metallicity and colour), to evaluate the parameter differences $\Delta_{i,j}$ between the star in question and the mean values of all evaluation subsets and smooth via the exponential kernel with the smoothing lengths $\sigma_{j}$. I chose $\sigma_{\feh} = 0.1 \dex$ and $\sigma_{(g-i)_0} = 0.04 \mag$. When binning the sample a second time for measuring Galactic parameters, I fit the distance for a second time to diminish the danger of remaining systematic biases. Anyway there is no evident trend, e.g. with distance, in the second step and the remaining corrections are on the noise level.

\cite{SBA} showed that in the presence of significant intrinsic distance scatter the method a slight mean distance underestimate occurs, which is caused by stars with distance overestimates populating the edges of the fitting baseline and hence acquiring larger weight. This bias was found at $\delta_f \sim 0.5 \sigma_{f}^2$, where $f$ denotes the relative distance error and $\sigma_f^2$ its variance. I confirmed the validity of the prefactor $0.5$ for the sample in use by applying a range of Gaussian broadenings to the first guess isochrone distances. Though the prefactor is known, it is not straight forward to estimate the values of $\sigma_f$ dependent on metallicity and colour. On the red side the stars are relatively firmly placed on the main sequence, the reddening vector runs quite parallel to the main sequence, and hence the colour and metallicity errors dominate justifying a moderate error estimate $\sigma_f \sim 15\%$. On the blue end the placement of the stars on the turn-off becomes increasingly uncertain and especially on the metal-rich side one must expect significant spread in ages and hence absolute magnitudes as well. I hence adopt a mild increase in distance depending on colour
\begin{equation}
\frac{\delta_s}{s} = 0.01 + 0.06\left({\rm(g-i)}_0-0.8\right)^2\left(1 + \frac{1}{4}\left(\feh+2.3\right)^2 \right), 
\end{equation}

which is larger on the blue side and increases mildly with metallicity. I limit the relative distance increase to a maximum of $4.5\%$, which corresponds to a distance scatter $\sigma_f = 30\%$.

\figref{fig:p7distmap} shows the finally adopted absolute magnitudes (upper panel) and the correction field for the isochrone distances (lower panel) in the colour-metallicity plane. Along the blue metal-poor stars I require moderate distance stretching. This is the turn--off region, where the $11.6 \Gyr$ isochrones might be a little bit too young and get shifted upwards in addition to the known suspicion of isochrones being too faint. On the other hand a bit of contamination by subgiants is more likely because of the reduced difference in surface gravity and weaker absorption lines. The problems with gravities for the most metal-poor stars are indicated by the absolute magnitudes bending back, i.e. the most metal-poor stars being made brighter again at the same colour. Carbon stars constitute another bias having stronger absorption in the g-band, making them too ``red'' in the colour assignment, which brings forth a brightness underestimate.

In principle the method explores the structure around the turn--off and the dominant ages among these populations, but owing to the scope of this work I defer this to a later paper. At higher metallicities the turn--off line evidently bends upwards, i.e. to redder colours as expected from stellar models. On the other hand there is evidence for a younger metal-rich population at blue colours, indicated by the absence of distance under-estimates contrasting with their metal-poor antipodes. Some vertically aligned features might point to minor systematics in the pipeline metallicities (e.g. under-estimating a star's metallicity makes us underestimate its distance). The little spike near the turn-off around $\feh \sim -1.0$ could be an unlucky statistical fluctuation. If it was real and not caused by a problem in the pipeline it would indicate a rather sharp transition in ages with relatively young stars at slightly higher metallicities opposing the very old population in the halo regime. Throughout the range of G and K dwarfs there is a quite constant need for mildly larger distances than envisaged by naive use of the stellar models.

To test my approach for consistency, I experimented with the \cite{Ivz08} (A7) relation as well and found that as expected the distance corrections robustly produce an almost identical final outcome despite the highly different input.

\subsubsection{Radial velocities and vertical motion of the Sun}\label{sec:p7rad}
Despite all efforts put into the SEGUE parameter pipeline, it cannot be granted that we can trust the radial velocities to be free of any systematic errors of order $1 \kms$. To shed light on this problem, let us examine the vertical velocity component. A good estimate of vertical motion can be derived from the Galactic polar cones, where proper motions and hence distance problems have almost no impact on the measurement. Luckily SEGUE has both the Galactic North and South pole and I use $|\sin{b}| > 0.8$, while checking that a stricter cut at $0.9$ does not significantly alter the findings.
These polar cones display a surprisingly low $\Wsun = (4.96 \pm 0.44) \kms$. This is $2 \kms$ off the old value from Hipparcos and the GCS. While this may of course trace back to some unnoticed stream in the solar neighbourhood, it seems appropriate to check for consistency. $14272$ stars towards the North Pole show $\Wsun = (4.26 \pm 0.52) \kms$, while $3482$ stars towards the South Pole give $\Wsun = (7.84 \pm 0.85)\kms$. Allowance for a variation in the line--of--sight velocities:
\begin{equation}
W = \delta v_{\rm los} \sin{b} - \Wsun 
\end{equation}
estimates $\delta v_{\rm los} = (2.04 \pm 0.63) \kms$ and $\Wsun = (-6.07 \pm 0.56) \kms$. The tighter cut in latitude gives the same line--of--sight velocity deviation, but $\Wsun = -6.9 \kms$. Binning in metallicity brings me to the limits of statistical significance, but reveals a rather robust uptrend in the estimated $v_{\rm los}$ offset from near to negligible at solar metallicity to of order $5 \kms$ near $[Fe/H] \sim -2$. It is tempting to argue in accordance with \cite{LawrenceYanny12} that all those trends are reaction of the Galactic disc to some satellite interaction, but such reasoning is contradicted by more of the signal coming from the halo stars, which cannot participate in such a response. An explanation by streams conflicts with the broad metallicity range concerned that would require a conspiracy of nature. A simpler explanation might be a warp in the disc, but again this conflicts with the metallicity trend.
The easiest explanation for the deviation is a residual error in the pipeline. To avail this and stay as close as possible to the pipeline values, I use henceforth a minimal correction that keeps the discussed trend just below significance:
\begin{equation}
v_{\rm los} = v_{\rm los, pip} + (-0.2 + 3.3\feh / 2.5)\kms
\end{equation}
allowing the corrections to vary between $-0.2$ and $-3.5 \kms$.

\section{Detecting global rotation in SEGUE}\label{sec:p7globrot}

\begin{figure}
\epsfig{file=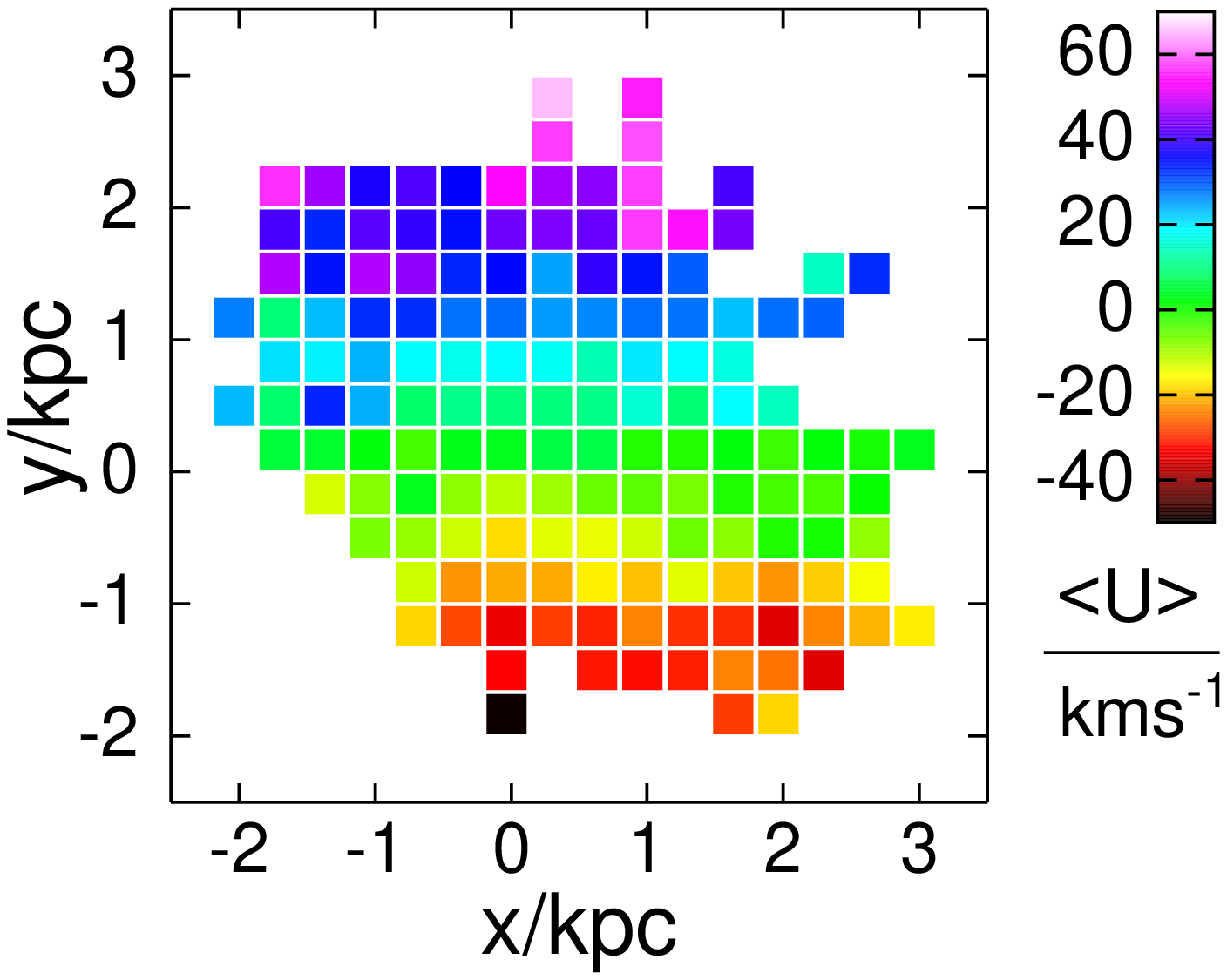,angle=-90,width=0.495\hsize}
\epsfig{file=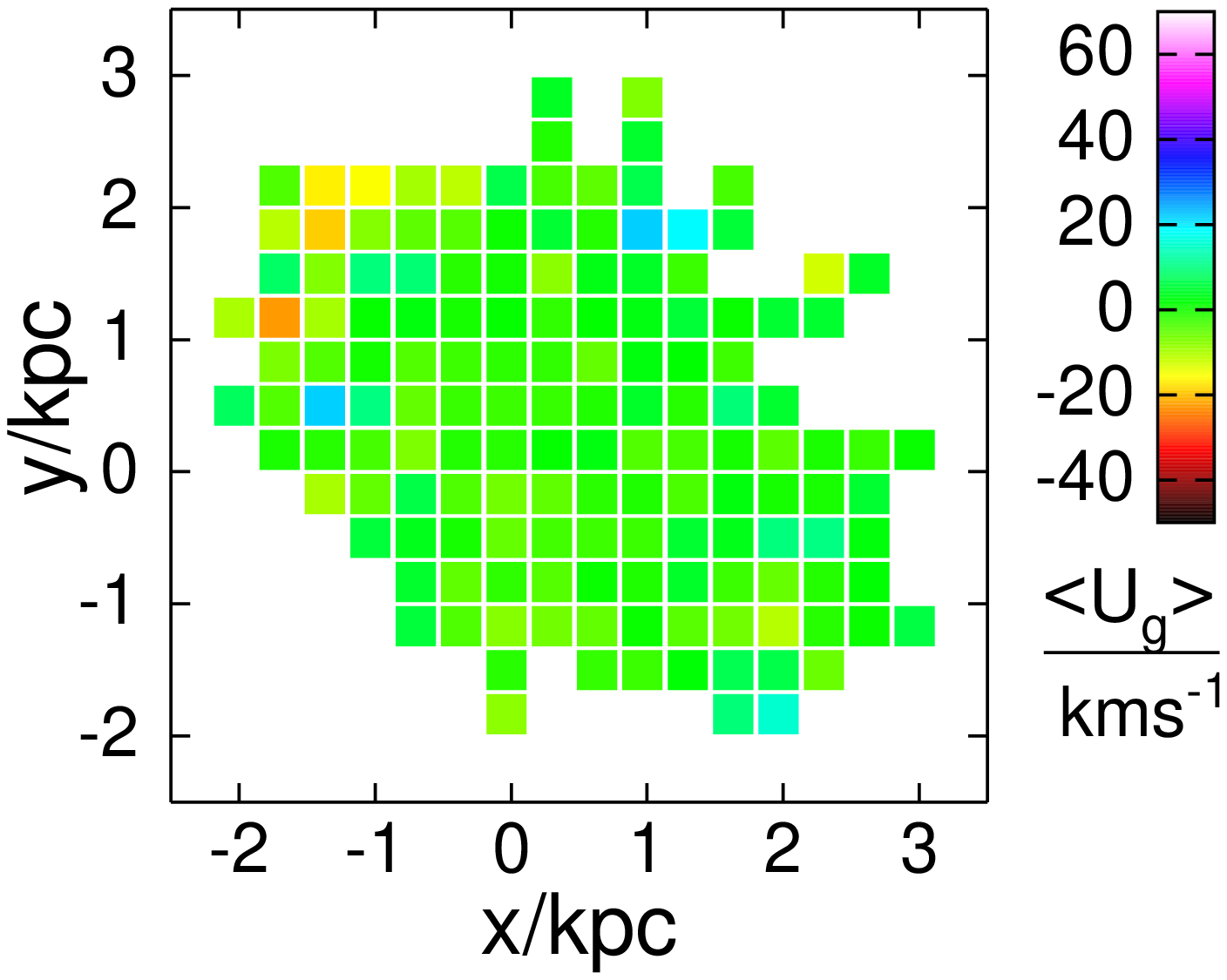,angle=-90,width=0.495\hsize}
\caption[Galactocentric and heliocentric radial velocities]{Systematic streaming in the SEGUE sample. The left hand panel shows the heliocentric $U$ velocity, while the right hand panel gives the galactocentric $U_g$, i.e. the real radial motion at the position of the star assuming a solar galactocentric radius $\Rsun = 8.2 \kpc$ and $\Vtot$ set accordingly to the proper motion of Sgr $A^*$. The sample was binned in separate boxes a third of a $\kpc$ wide, plotted when they contain more than $50$ stars. The sample thins out towards the remote stars, which causes the increased scatter seen particularly in the galactocentric velocities. For both panels I restrict the sample to stars with $\feh > -1$.
}\label{fig:p7Umap}
\end{figure}

\figref{fig:p7Umap} shows the systematic motion for the sample projected into and binned in the Galactic plane. At each bin I evaluate the mean motion in the heliocentric frame (left panel) and the mean motion in the Galactocentric frame (right). While I would expect some minor systematic motion even in the Galactocentric frame by spiral structure and maybe influence by the bar, such contributions will be relatively small as most of this sample is at high altitudes, where the stars with their large random velocities experience less important changes of motion by the relatively shallow potential troughs of the spiral pattern. Some of the residual structure will likely be blurred out by Poisson noise and distance uncertainties, so that it is no surprise not to see any appreciable signal. As a good sign there are no direction-dependent shifts, which would be expected under systematic distance errors, which make especially the azimuthal velocity component cross over into the radial velocity determinations. In the left panel, however, we see the very prominent rotational streaming of the Galactic disc, which is central to the following investigations.

\subsection{The Standard of Rest revisited}\label{sec:p7LSR}

The strong influence of rotational streaming on the observed radial motion makes any determination unreliable that does not explicitly take it into account. Before I turn to its positive application, I lay out a simple procedure to deal with this effect as a contaminant and measure the solar radial LSR velocity.
In the literature varying offsets of the solar radial motion of up to more than $10 \kms$ from the value derived from local samples have been claimed. Besides possible velocity cross-overs, e.g. the motion in $V$ translating into a vertical motion \citep[for an example see Fig. 7 in][]{SAC} invoked by systematic distance errors \citep[][]{SBA}, neglected sample rotation can be the reason. In any sample that is not demonstrably rotation-free and is not symmetric in Galactic longitude the streaming motion offsets the mean heliocentric radial motion and hence the inferred Solar motion. The error grows with the remoteness of the observed stars.

\begin{figure}
\epsfig{file=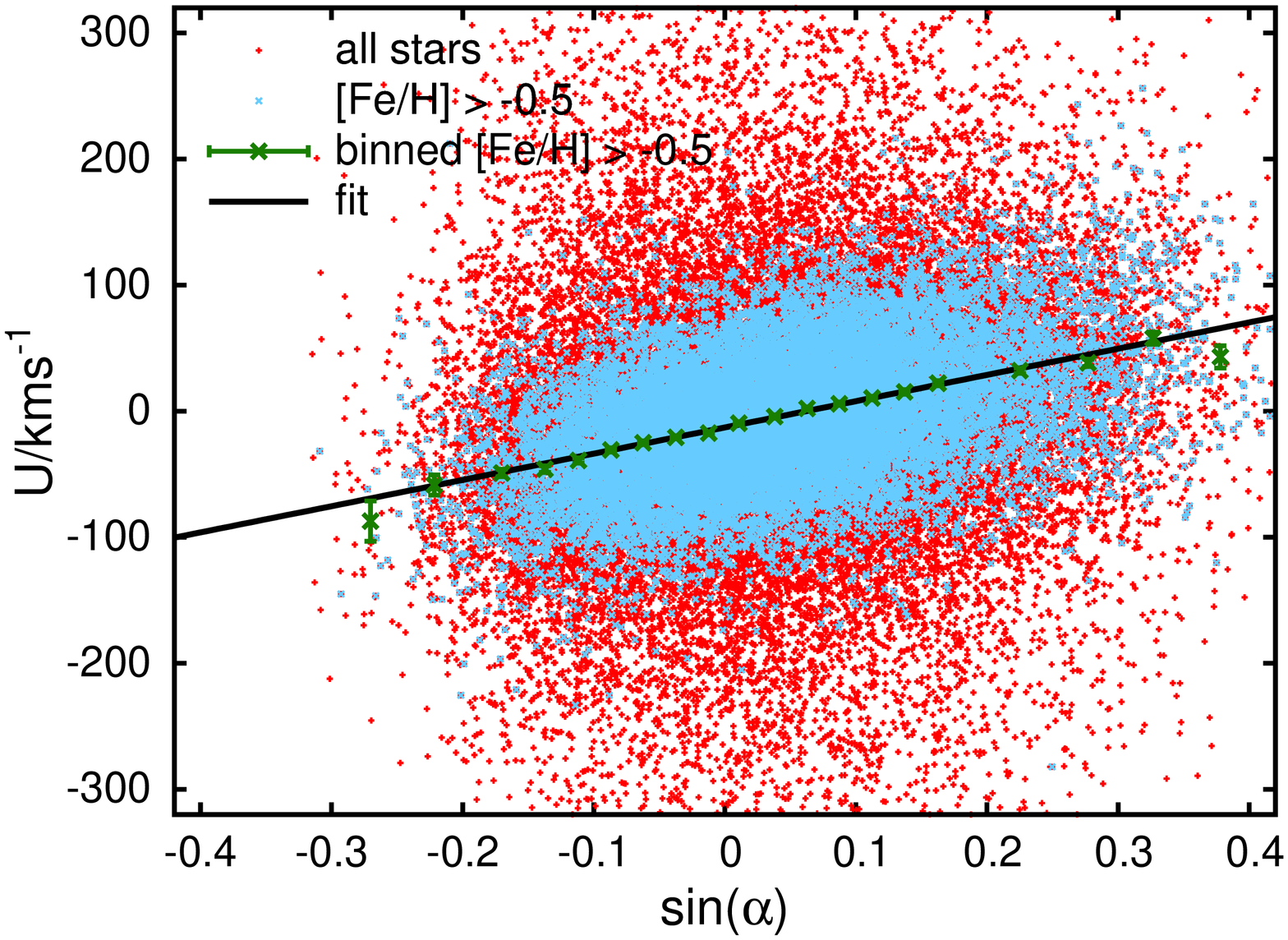,angle=-90,width=\hsize}
\epsfig{file=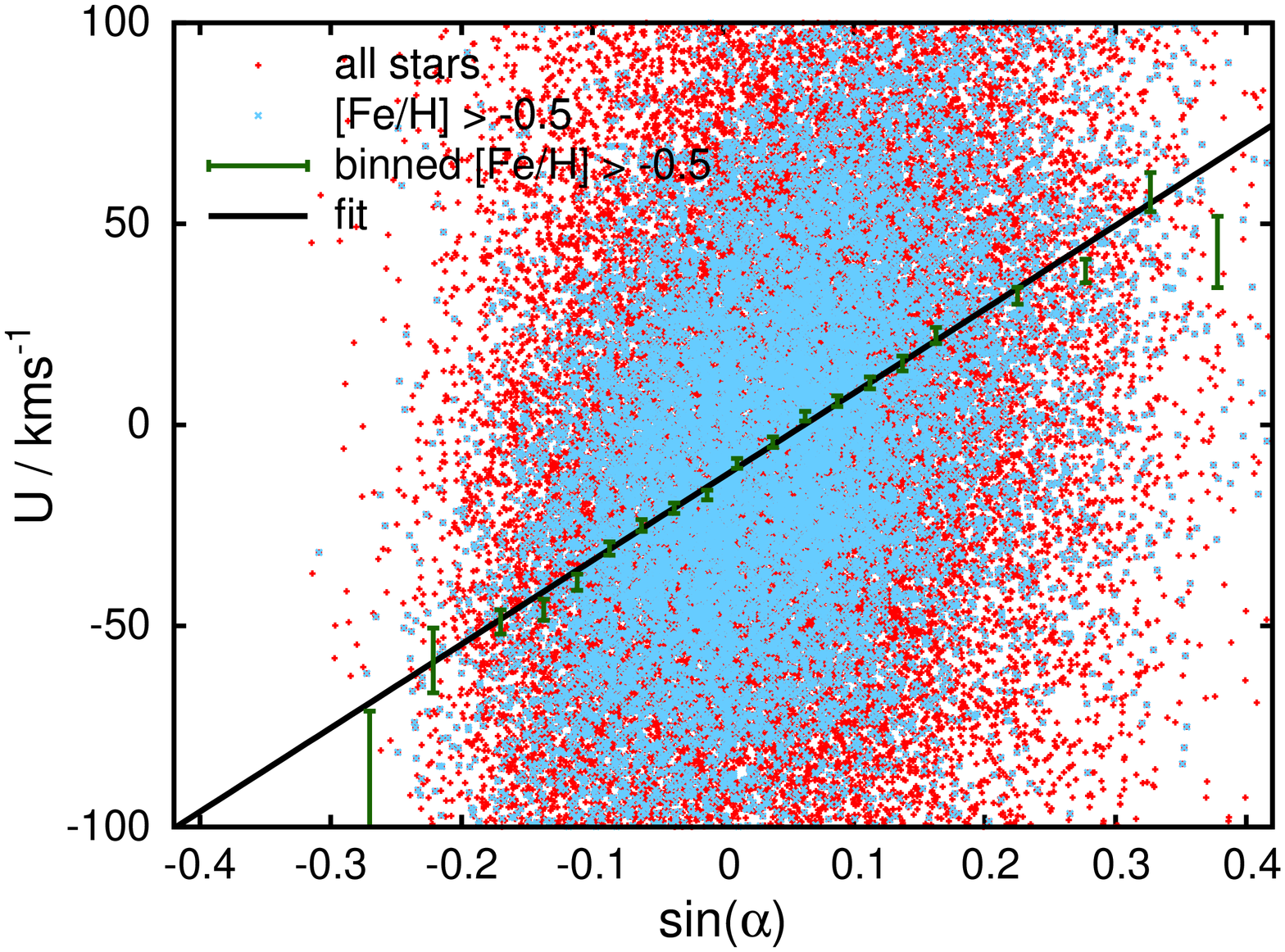,angle=-90,width=\hsize}
\caption[Heliocentric $U$ velocities versus $\sin{\alpha}$]{$U$ velocities versus $\sin{\alpha}$ for all stars in the SEGUE sample (red crosses) and objects at disc metallicities ($\feh > -0.5$, blue points). The line represents the linear fit for the metal-rich subsample, the dark green error bars indicate the mean velocities in bins of $\sin{\alpha}$. The lower panel is a close-up of the central region. 
}\label{fig:p7Utrend}
\end{figure}

\figref{fig:p7Utrend} shows the heliocentric $U$ velocities for all stars (red crosses) from DR8 and stars with $\feh > -0.5$ (light blue points) against $\sin \alpha$. I set $\Rsun = 8.2 \kpc$, and the value for $\Vtot$ accordingly to $248.5 \kms$. The metal-rich subsample displays a nearly linear relationship of the mean $U$ velocity versus $\sin \alpha$ as expected for rotational streaming from equation (\ref{eq:p7rot}). This streaming motion is far larger than the expected offset caused by the solar LSR motion $\Usun \sim 11.1 \kms$ \citep[][]{SBD}. Even the mild asymmetry in $\alpha$ distorts a naive estimate of $\Usun$ in disregard of rotation to $\Usun = (3.85 \pm 0.43) \kms$ for metal-rich objects with $\feh > -0.5$ and $\Usun = (6.77 \pm 0.44) \kms$ for the entire sample. Consistent with the rotational bias the deviation from the standard LSR value is significantly larger for the disc sample. To rid my results from this bias, I use the estimator
\begin{equation}\label{eq:p7Ufit}
U(\alpha) = \theta \sin{\alpha} - \Usun + \delta_U
\end{equation}
where the free parameters $\theta$ and $-\Usun$ are the effective rotation and the reflex of the solar motion, $\delta_U$ describes noise from both velocity dispersion and observational scatter that should have zero mean independent of $\alpha$. I obtain $(\theta,\Usun) = (208.2 \pm 3.6, 12.90 \pm 0.42)\kms$ for the metal-rich subsample and $(\theta, \Usun) = (149.3 \pm 3.7, 13.04 \pm 0.46)\kms$ for all stars. The formal errors are mildly underestimated due to the non-Gaussianity of the $U$ velocity distribution and a non-weighted fit. This differs from the \cite{SBD} estimate at marginal significance. 

Yet there are larger uncontrolled systematic biases: This fit implies that the effective rotation term carries away all the observed bias on $\Usun$. While the detailed rotation pattern is not of interest here, the measurement will still be biased if one encounters a larger fraction of slow rotators on one side than on the other. E.g. a different number of halo stars or unmatched differences of rotation with galactocentric radius can distort $\theta$, which correlates with $\Usun$. In this sample $\Usun$ shrinks by $0.5 \kms$ for every $10 \kms$ decrease in $\theta$. In the present case such effects are below detectability, as can be seen from the lower panel in \figref{fig:p7Utrend}, where the binned means (green error bars) show no significant deviation from the linear relationship.
Further, any remaining systematic distance errors can lead to systematic velocity shifts. The first error can be controlled by selecting only stars with small $\alpha$, the second by selecting stars near the Galactic centre and anticentre directions. In principle this could be tested by geometric cuts on the sample, however, there are no data near the Galactic plane. Exploring the full parameter space of possible cuts reveals some expected fluctuations of order $1.5 \kms$, but no striking systematics. For example $|\sin{\alpha}| < 0.1$ gives $\Usun = (13.29 \pm 0.54) \kms$ in concordance with the full sample.

More importantly, cuts for a smaller distance range, which is more reliable concerning the proper motion uncertainties favour $\Usun \sim 14\kms$. The difference is caused by the $\sim 15 \%$ most remote stars displaying a rather extreme $\Usun \sim 9 \kms$. E.g. cutting for $s < 2.5 \kpc$ I get $\Usun = (14.26 \pm 0.50) \kms$, again consistent with $\Usun = (14.08 \pm 0.57) \kms$ for $|\sin{\alpha}| < 0.1$, demonstrating that this drift in distance is not an effect of some rotation error. A similar rise in $\Usun$ to up to $15 \kms$ can be achieved by increasingly tight cuts on the signal to noise ratio, which indirectly selects a shorter mean distance. A global line-of-sight velocity increment $\delta v_{\rm los} = 2 \kms$ increases the estimate for $\Usun$ by about $0.5 \kms$. 

It is interesting to investigate into the question of a trend in radial velocities with galactocentric radius $R$ as suggested in \cite{Siebert11}. To this purpose I estimate the equation
\begin{equation}
U = \theta \sin\alpha + \eta (R-\Rsun) - \Usun + \epsilon_U 
\end{equation}
where $\theta$ is the effective rotation, $R$ the galactocentric radius, $\eta = d\langle U \rangle (R) / dR$ is the effective (heliocentric) radial velocity gradient on the sample, and $\epsilon_U$ is the noise term from velocity dispersion. For convenience I limit $|\sin\alpha| < 0.1$ as before and get on the full distance range $\eta = (1.53 \pm 0.55) \kms\kpc^{-1}$ at $\Usun = (14.22 \pm 0.64) \kms$. On the shorter distance range of $s < 2.5$ I get $\Usun = (14.22 \pm 0.67) \kms$ and $\eta = (0.29 \pm 0.7) \kms\kpc^{-1}$, delivering no sensible gradient at all. Restricting $\feh > -0.5$ I get $\eta = (1.05 \pm 0.7) \kms\kpc^{-1}$. This residual trend is caused by the $1000$ stars with $\feh > 0$ measuring $\eta = (3.7 \pm 1.9) \kms\kpc^{-1}$. While this may finally be the desired signal, it is puzzling to have no more metal-poor counterpart. Hence this is more likely either be a statistical fluctuation or some flaw in the sample. Parameters and distances for stars with super-solar metallicity in SEGUE are likely problematic, particularly as I do not have sufficient numbers to constrain their mean distance very well. It has to be noted that this sample is at rather high altitudes and is dominantly in the outer disc, so that velocity trends in objects of the inner thin disc might not show up. Hence the missing detection of an appreciable trend (and exclusion of a significant deviation from circularity) does not disprove \cite{Siebert11}. More of a problem for their result should be the larger value of $\Usun$, which likely reduces or diminishes their estimated gradient, as their analysis mainly relies on assumption of the old value and does not have sufficient data points in the outer disc to provide a decent fix.

After all, when accounting for the systematic uncertainties in the data, I cannot detect any reliable hint for a motion of the solar neighbourhood against some more global Galactic Standard of Rest. Combining these estimates with the result from the mean direction of motion (see Section \ref{sec:p7direct}) of $(13.9 \pm 0.3) \kms$ with a systematic uncertainty of $\sim 1.5 \kms$ and the reader may either use this value or a lower $\Usun = (13.0 \pm 1.5) \kms$, taking into account the value to the older GCS.

\begin{figure}
\epsfig{file=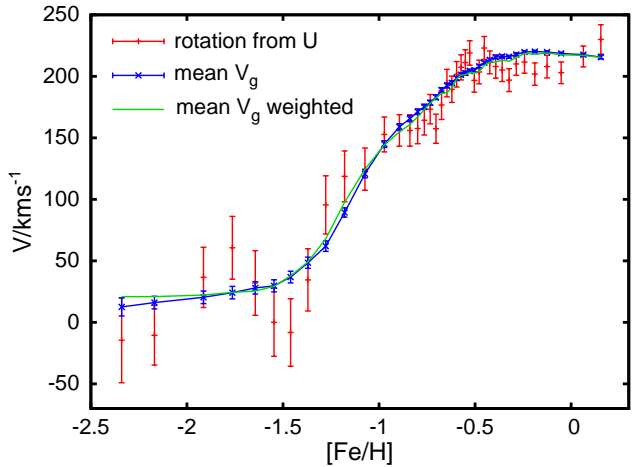,angle=-90,width=\hsize}
\caption[Measuring the intrinsic rotation of subpopulations in metallicity]{Measuring the intrinsic rotation of the populations. I assume $\Rsun = 8.2 \kpc$ and $\Vtot = 248.5 \kms$. After sorting the sample in $\feh$ I move a $\sim 2500$ stars wide mask in steps of $1250$ so that every second data point is independent. The red data points show the measured rotation speed in each subsample, while the blue data points with small errors give the mean azimuthal velocity $\langle V_g \rangle$ in each bin. The green line shows $\left\langle{V_g}\right\rangle_w = \sum_i w_i V_{g,i}$ from equation (\ref{eq:p7weightedVg}), which is weighted with $(\sin{\alpha} - \overline{\sin{\alpha}})^2$.
}\label{fig:p7rotmet}
\end{figure}

\begin{figure}
\epsfig{file=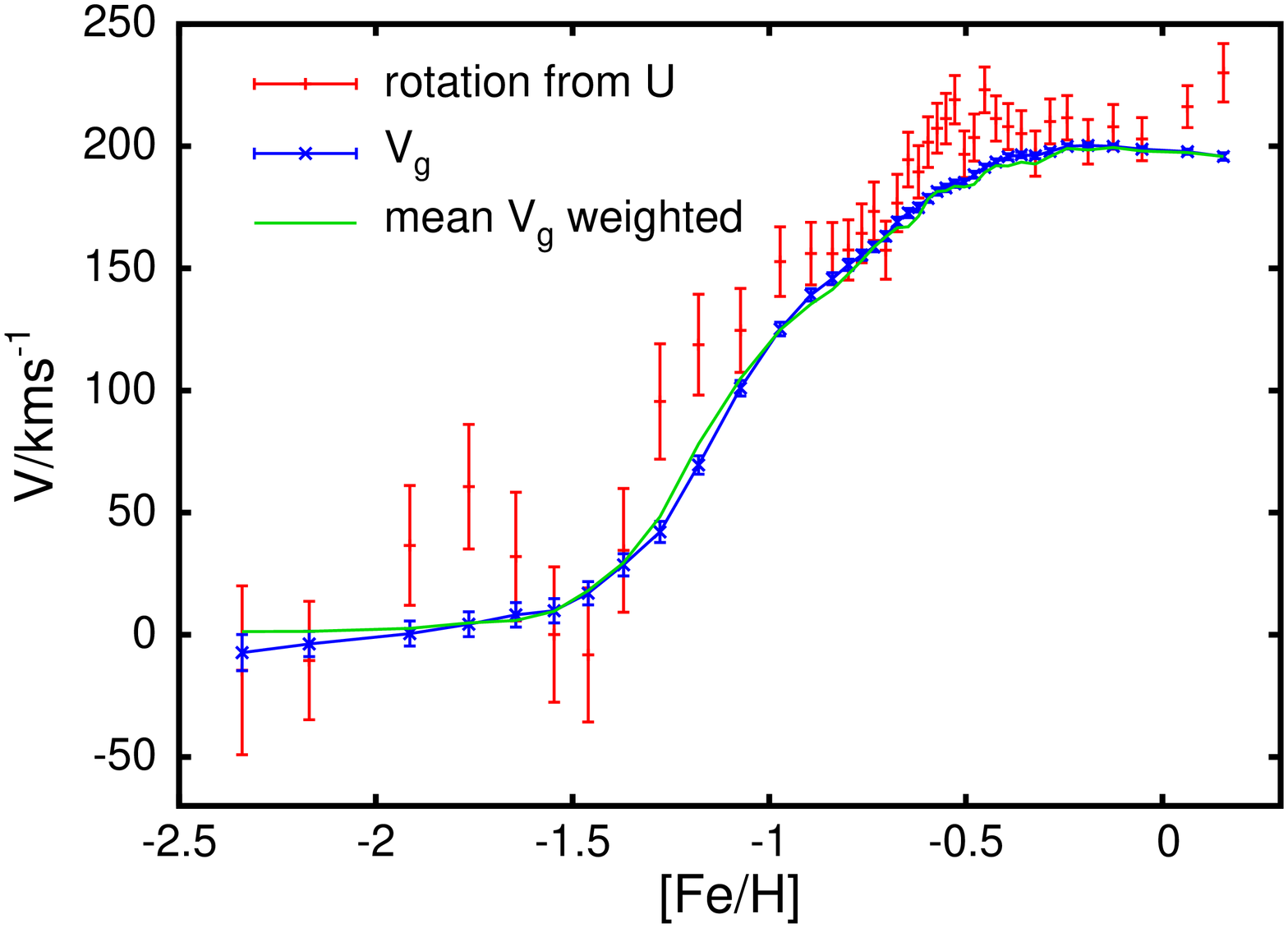,angle=-90,width=\hsize}
\epsfig{file=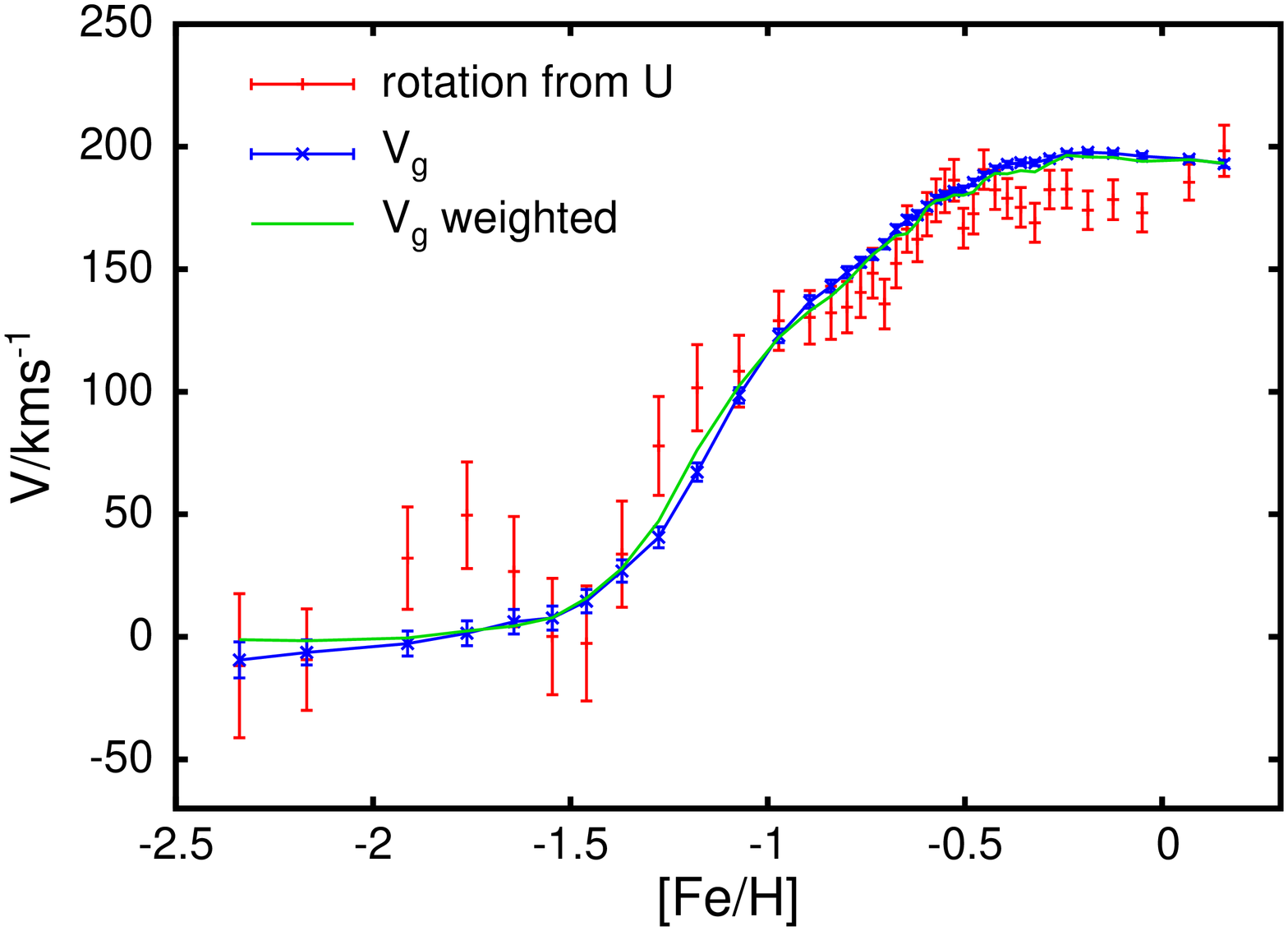,angle=-90,width=\hsize}
\caption[Effects of wrong circular speed and galactocentric distance]{Demonstrating the detectable effect of errors in Galactic parameters. Settings are identical to \figref{fig:p7rotmet}, with one difference each. In the top panel I set $\Vtot = 228.5 \kms$ close to the IAU standard value. While the measured rotation speed from $U$ stays identical, the determination by $V_g$ is shifted down into stark disagreement. In the bottom panel I set $\Rsun = 7.0 \kpc$ and lower the Solar motion to $\Vtot = 225.5 \kms$. The compression in the measurement from $U$ velocities compared to the $V_g$ measurement is clearly visible.
}\label{fig:p7rotmetf}
\end{figure}

\subsection{Divide and conquer: The rotation of components}\label{sec:p7metrot}

So far I exclusively looked at the global rotation pattern and its influence on $\Usun$, but the new rotation estimator can do more: Any measurement relying on observed azimuthal velocities does not provide the real rotation of the component in question, but only a value relative to the Sun, from which we conclude its azimuthal speed by discounting for the Galactic circular velocity plus the solar azimuthal LSR motion (plus a small part of the solar radial motion when $|\alpha| > 0$). 
On the contrary the radial velocity part of (\ref{eq:p7rot}) uses the velocity differences at a range of Galactic positions and gives the ``real'' rotation of the component without significant influence by assumptions on the motion of the Sun. The measurement is in most cases less precise than the azimuthal measurement, but is nevertheless interesting by its different set of assumptions. 
Consider a possible presence of stellar streams: They can distort azimuthal velocity measurements soliciting an independent source of information: In general a stream will then show up in highly different values for the two rotational velocity indicators. The price one has to pay for the radial velocity estimator is a larger formal error, the need for extended samples, a different dependence on the assumed $\Rsun$ (which is usually a bonus) and some stronger vulnerability to distortions from axisymmetry, e.g. influence by the Galactic bar.

As discussed by \cite{SBA}, dissecting a sample by kinematics is detrimental for this kind of kinematic analysis, but one can divide the Galaxy into its components by a selection in metallicity.

\figref{fig:p7rotmet} shows the rotation measurement from the radial velocity estimator when I dissect the sample into bins of $\sim 2500$ objects in metallicity and move half a bin in each step (large red error bars). The two plateaus of disc and halo rotation with the transition region in between show up nicely. The plateau of disc rotation stays about $15\kms$ below the circular speed of the disc due to asymmetric drift. Clearly there is no significant average net rotation in the metal-poor Galactic halo. Despite the far smaller formal error, this would be more difficult to state on firm grounds by using the average azimuthal velocity alone (blue error bars), which does depend strongly on the solar Galactocentric motion. In fact the azimuthal velocity estimator has been shifted onto the $U$ estimator's curve by matching $\Vtot = 248.5 \kms$, which turns us to the topic of the next subsection.

\subsection{Deriving Galactic parameters}\label{sec:p7Galparam}

More important than mutual confirmation of results one can play off the two rotation measurements (from $U$ and $V_g$) against each other. As stated before, only $V_g$ depends on the solar azimuthal velocity, so that forcing both rotation estimators into agreement we obtain an estimate for the total solar azimuthal velocity in the Galactic rest frame $\Vtot$. By subtraction of the solar LSR motion this directly estimates the Galactic circular speed $V_c$ near the solar radius without any need to invoke complicated models. The apparent difference caused by a wrong $\Vtot$ is demonstrated in the top panel of \figref{fig:p7rotmetf}. For this plot I lowered $\Vtot$ from the optimal value by about $20 \kms$. Comparison with the previous figure reveals that the red bars are unchanged as expected, while the azimuthal velocity based measurements are pushed upwards into stark disagreement.

To bring the two rotation measurements on common ground one has to apply weights to the estimate of the mean azimuthal velocity $\left\langle{V_g}\right\rangle$ for each subpopulation: In a least squares estimate on the rotation speed $\theta$ each star appropriates a weight proportional to its distance from the population mean on the baseline $d_{\alpha} = |\sin{\alpha} - \left\langle\sin{\alpha}\right\rangle|$. As the value $U_h$ invoked by rotation varies linearly on this baseline, each star attains the weight $w = (\sin{\alpha} - \left\langle\sin{\alpha}\right\rangle)^2$ when measuring rotation. Introducing a simultaneously weighted mean for the mean azimuthal velocity 
\begin{equation}\label{eq:p7weightedVg}
\left\langle{V_g}\right\rangle_w = \sum_i w_i V_{g,i} $,$
\end{equation}
where the sum runs over all stars $i$, any structural differences between the two estimators are eliminated, since every population enters both estimators with equal weights. This measurement of the Solar velocity does thus not require any modelling of Galactic populations or any other deeper understanding at all. The procedure is then very simple: 
\begin{itemize}
\item Assume $\Rsun$ and a first guess solar velocity vector $\Usun, \Vtot', \Wsun$
 \item Estimate $\theta$ and $\left\langle{V_g}\right\rangle_w$.
\item Reduce $\Vtot'$ by the difference $\left\langle{V_g}\right\rangle_w - \theta$ and recalculate $\theta$, $\left\langle{V_g}\right\rangle_w$ until convergence.
\end{itemize}
 
The only remaining uncertainty (apart from systematic measurement errors) can arise if the geometry of the sample allows for a distortion in the radial velocity estimator. I hence recommend controlling the value of $\Usun$, which is well-behaved for this sample, and in other case fixing it to a predetermined value. In the latter case the weights of $\left\langle{V_g}\right\rangle_w$ must be changed to $w = (\sin{\alpha})^2$. 

The weighted averages on the azimuthal velocities are drawn with the green line in \figref{fig:p7rotmet}. Generally the weighted values in the disc regime are a bit lower, because the extreme ends on $\sin{\alpha}$ are populated by remote stars at higher altitudes and at larger asymmetric drift outweighing other effects like the polewards lines of sight reaching higher altitudes.

\begin{figure}
\epsfig{file=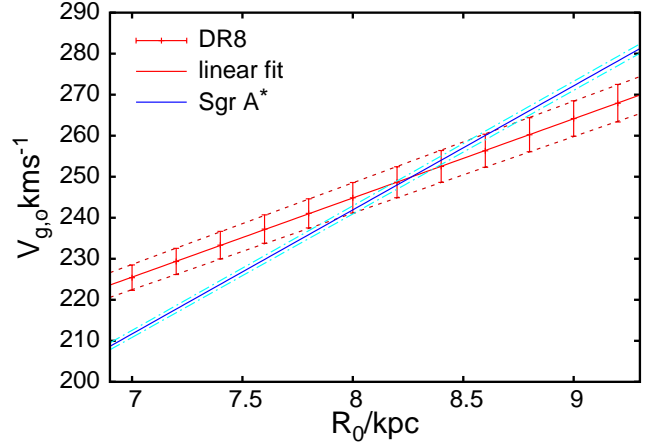,angle=-90,width=\hsize}
\caption[Measuring the total azimuthal velocity $\Vtot = V_c + \Vsun$ of the Sun]{Measuring the total azimuthal velocity $\Vtot = V_c + \Vsun$ of the Sun by forcing an agreement between the radial velocity based rotation estimator and the azimuthal velocities. For each assumed $\Rsun$ the sample was binned in distance at bins of $\sim 2500$ stars and then a weighted average taken, which is shown with red error bars. To guide the eye, I add the lines of a linear fit. For comparison I plot the constraint from the proper motion of Sgr $A^{*}$ with blue lines.
}\label{fig:p7distV}
\end{figure}

Via the Galactic angle $\alpha$ the assumed solar galactocentric radius enters the derivation. As depicted in \figref{fig:p7varr} the larger $\Rsun$, the smaller the derived $|\alpha|$ will be for stars on our side of the Galaxy. This results in a larger rotation speed $\theta$ required for the same observed pattern and so finally increases the estimate of $\Vtot$. Fortunately the rotation speed of Galactic components is lower than the solar value and further -- in contrast to the linear Oort constants -- the dependence on $\alpha$ is not linear in distance, which helps towards a relationship flatter than a mere proportionality between $V_c$ and $\Vsun$. This is shown with the red error bars in \figref{fig:p7distV}. For the practical calculation I bin the sample in distance and move a mask of $\sim 2500$ stars width in steps of $625$ stars. This stabilises the results against scatter from the coarse binning, the error is corrected for the induced autocorrelation. The result is then drawn from a weighted average of the distance bins. The binning in distance is mandatory in this sample not mainly because of reddening uncertainties, but because of the previously described anomaly in the mean radial velocities with distance. This way contaminations are kept at a minimum level.

By using the proper motion of Sgr $A^*$ \citep[determined as $6.379 \pm 0.026 \mas \yr^{-1}$ by][]{Reid04} as a second constraint the Galactocentric radius can be fixed to $\Rsun = 8.26^{+0.37}_{-0.33} \kpc$ and $\Vtot = 250^{+11}_{-10} \kms$. For the resulting circular speed one has to subtract about $12 \kms$. 

\begin{figure}
\epsfig{file=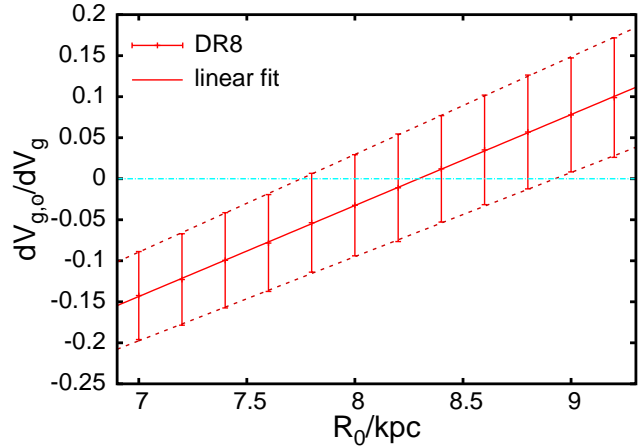,angle=-90,width=\hsize}
\caption[Using the trend in determined rotation speed versus mean azimuthal velocity to get another constraint to the galactocentric radius]{Using the trend in determined rotation speed versus mean azimuthal velocity to get another constraint to the galactocentric radius. For this purpose I sorted the sample by metallicity into bins of $2500$ stars and estimated $d\Vtot/dV_g$ (vertical axis in the plot,a dimensionless quantity) while varying $\Rsun$.
}\label{fig:p7Vctrend}
\end{figure}

One can even gain an estimate for the Galactocentric radius that is fully independent of any other results. For this purpose we have to recapitulate again the impact of $\Rsun$ on the derived $\Vtot$ for different populations, shown in \figref{fig:p7varr}: The measurement of $\Vtot$ compares the weighted azimuthal velocity in any population to the derived absolute rotation $\theta$. Changing the Galactocentric radius $\Rsun$ makes the Galactic coordinate system bend differently and hence gives only minor changes to $V_g$, but strongly affects $\sin{\alpha}$. This way $\theta$ is nearly proportional to $\Rsun$. A weakly rotating population has a far smaller value of $\theta$ and hence experiences a far smaller change than fast rotating disc stars. If we now underestimate $\Rsun$, the estimates for $\theta$ are compressed compared to the $V_g$ based estimator as demonstrated in the bottom panel of \figref{fig:p7rotmetf}. In other words the estimates of $\Vtot$ for disc stars get sheared against the estimates for halo stars. However, for all populations one measures the same quantity, i.e. the azimuthal speed of the Sun, which cannot change between the samples. A deviation in $\Rsun$ is hence easily detected by dissecting the sample in metallicity, which is (as can be seen from Figure ~\ref{fig:p7rotmet}) a very good proxy for rotation. I cut the sample again into bins of $\sim 2500$ stars -- now in metallicity instead of distance -- and evaluate the trend $\Vtot$ $d\Vtot/d{\left\langle V_g\right\rangle}_w$ of $\Vtot$ against the mean azimuthal speed between the bins by fitting the linear equation $\Vtot = (d\Vtot/d{\left\langle V_g\right\rangle}_w) \left\langle V_g\right\rangle + V_0$ with $V_0$ as additional free parameter. From the red error bars in \figref{fig:p7Vctrend} one can see that already on the quite modest sample size and extension of the $\sim 50000$ usable stars in DR8 I get a useful, independent estimate of the solar Galactocentric radius $\Rsun = 8.29^{+0.63}_{-0.54} \kpc$. 

As in the first measurement I used the absolute value of rotation, while here the unit-less slope $d\Vtot/d{\left\langle V_g\right\rangle}_w$ is employed, the two results are formally independent and can be combined to $\Rsun = (8.27 \pm 0.29) \kpc$ in excellent agreement with \cite{McMillan11}. This translates to $\Vtot = (250 \pm 9) \kms$. Working in the LSR value of \cite{SBD} I hence obtain $V_c = (238 \pm 9) \kms$.

\subsection{Assessment of systematic errors}\label{sec:p7systerr}

To assess the systematic uncertainties I vary some of the assumptions. First I checked that the choice of bin size does not change the results more than the general noise induced by bin changes as long as the bins are not unreasonably small. Galactic rotation at fixed $\Rsun$ is increased by about $2 \kms$, when I globally augment distances by $10 \%$ and would be about $4 \kms$ lower without the \cite{SBA} distance correction. Increasing the distance scatter by applying an additional $10 \%$ raises $\Vtot$ by $0.6 \kms$ at fixed radius, application of a $30 \%$ error lowers it by $1.4 \kms$. The mild lowering is expected, since the strong distance scatter should give a global distance underestimate of $4.5 \%$ (consistently the rotation measurement from $U$ velocities drops by $\sim 2 \kms$). It is obvious that such a scatter has a minor effect, as the relation between $U_h$ and $\sin{\alpha}$ is linear and hence at small changes the mean of $U_h$ is only mildly affected. This is a benefit from operating remote from the Galactic centre where the changes in Galactic angle become more rapid. The mean azimuthal velocity is mostly affected by distance scatter via the bias on average distance derived from our statistical method. Tests reveal an expected systematic bias of order $1 \kms$. 

Similarly the slope of the derived rotation speed $d\Vtot / d V_g$ depends mostly on distance. Here the dependence is larger: It reacts by $0.08$, i.e. about $1.5$ standard deviations to a forced increase of $10 \%$ in distance and by $-0.08$ to a forced decrease of $10 \%$. Removing the statistical distance correction would lift values by about $0.2$. Increased distance scatter without distance correction increases the value by $0.011$ for an additional scatter of $10\%$ and lowers it by $0.023$ for a scatter of $30 \%$. 

Gravity cuts have minor influence on my results: The mean rotation rate is lowered by $0.9 \kms$ when I lift the minimum gravity to $\llg > 4.1$ and at the same time the trend in azimuthal velocity rises by $0.02$. Tighter cuts are not feasible for the trend, as I lose all metal poor stars. Replacing the sloping cut by a fixed $\llg > 4.0$, did not significantly alter the results ether. While the distance estimator strongly hints to some contamination of the dwarf sample, the observed deviation in $\Rsun$ is still not significant and likely results partly from random scatter, as many metal-poor objects are lost by this tighter cut. The robustness of the result against changes in the gravity selection points to the statistical distance estimator being able to cope with the different contamination by adjusting the mean distances. 

The largest uncertainty appears to arise from the proper motion determinations. The correction I apply raises rotational velocities by about $6 \kms$ and in light of the unsatisfactory physical reasoning as discussed in Section \ref{sec:p7pm} it is uncertain how well the correction works. A little reassurance is given by the second estimator, its unitless slope rises by only $0.01$ upon removal of the proper motion correction. The applied line-of-sight velocity correction lowers $\Vtot$ by about $1.2 \kms$ and lowers the unitless slope by just $-0.008$. The better robustness of the second estimator for $\Rsun$ reflects the fact that any spurious rotation affects both halo and disc rotation in a similar way and hence their difference is more robust than the mean. 

Further I find that a change of $\Usun$ in the calculation of galactocentric azimuthal velocities (a minor bias is expected in the lopsided sample) is irrelevant to the estimated quantities. However, if the rotation estimates themselves are done with fixed $\Usun$ instead of using it as a free parameter, the estimates agree fully when setting $\Usun = 13.9 \kms$, and $\Vtot$ falls by about $4 \kms$ for every $1 \kms$ reduction in $\Usun$. Depending on how much integrity we ascribe to the astrometry in the sample this sets an interesting relation between $\Usun$ and $\Rsun$. Similarly $\Usun$ can be determined by demanding the estimate for $\Vtot$ to be consistent in Galactic longitude $l$. Not surprisingly the favoured value is again $\Usun \sim 14 \kms$. I further tested that there are no trends of the rotation estimate (with fixed $\Usun$) with colour, distance or latitude.

The estimator of $\Usun$ from Section \ref{sec:p7LSR} depends weakly on distance. $\Usun$ increases by about $0.8 \kms$ for a global $10\%$ increase in distances and decreases by about the same amount for a distance increase. Its counterpart $\theta$ changes by around $3 \kms$ with some fluctuations as different distance cuts can affect the sample composition. The proper motion correction is responsible for about $6 \kms$ of the measured $\theta$ and $0.4 \kms$ of $\Usun$, while the line-of-sight velocity correction acts in the opposite direction, lowering $\theta$ by $2 \kms$ and $\Usun$ by $\sim 0.2 \kms$. 

\begin{figure}
\begin{center}
\epsfig{file=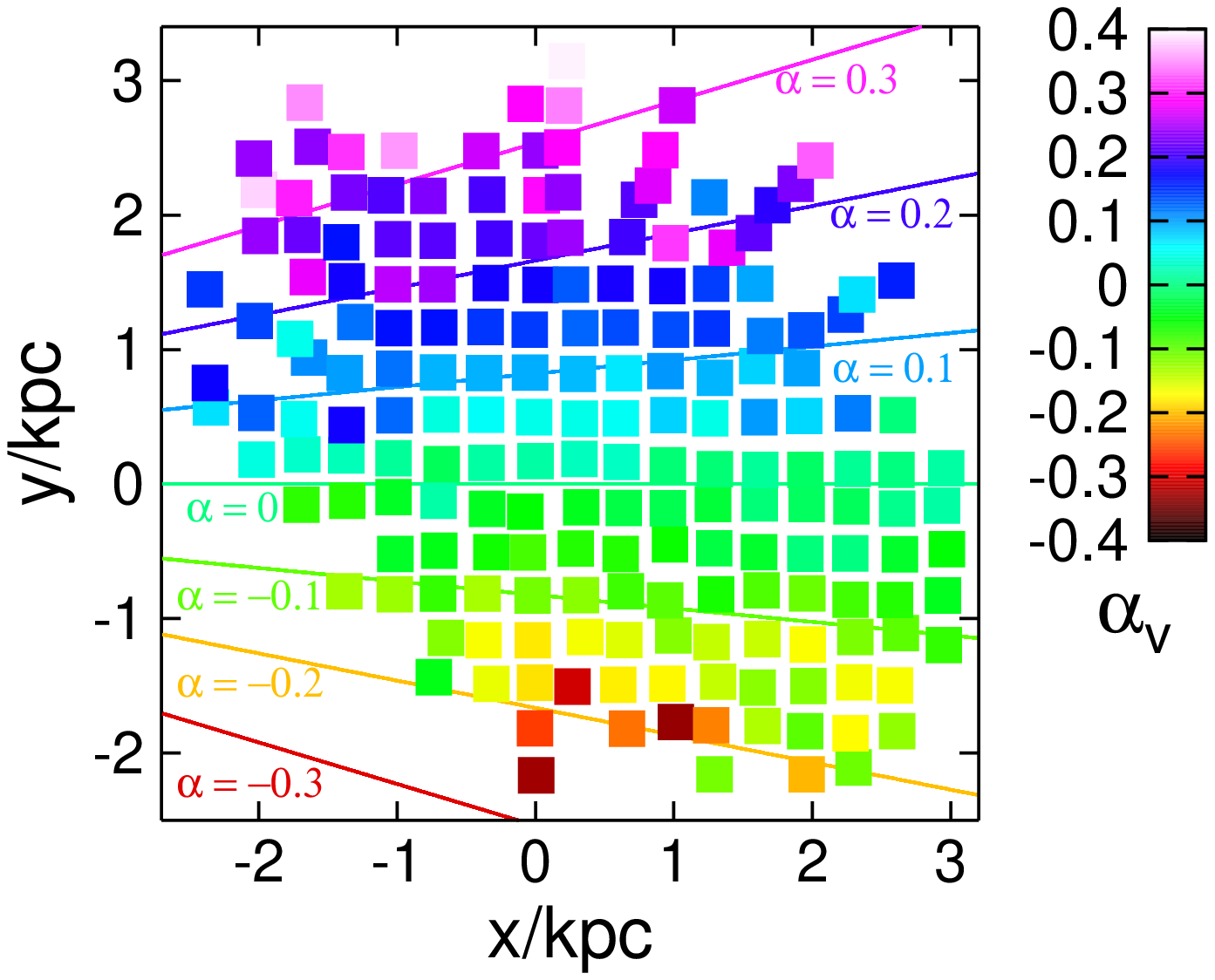,angle=-90,width=\hsize}
\epsfig{file=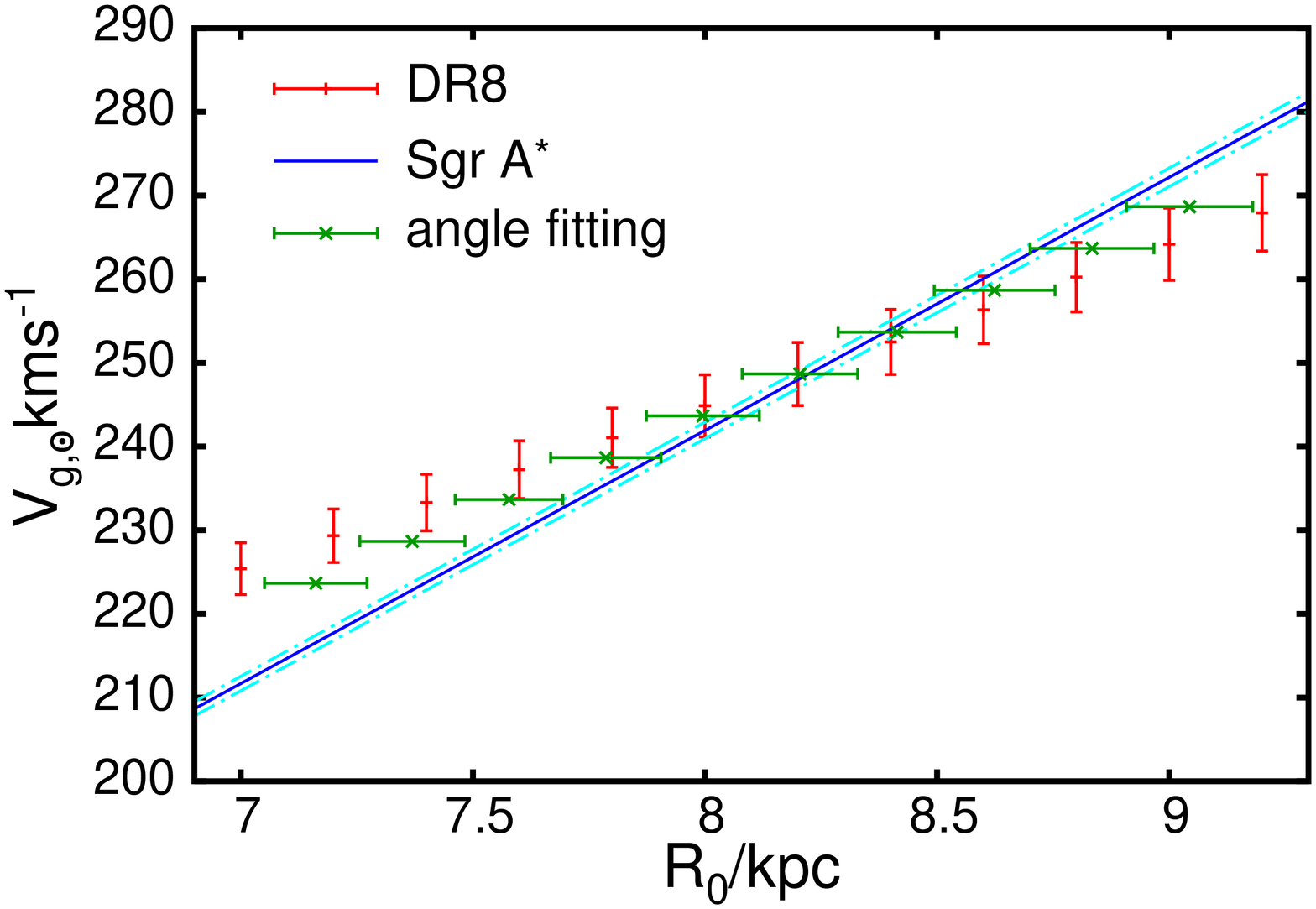,angle=-90,width=\hsize}
\end{center}
\caption[Mean motion direction in the disc plane and resulting estimates for $\Rsun$ versus adopted $\Vtot$]{The upper panel shows the angle $\alpha_v$ from equation (\ref{eq:p7alphav}) against the position angle $\alpha$ (in rad) of the mean motion vector in the plane for bins with more than $10$ stars. The distribution of bins is irregular because the mean values in $x$ and $y$ were used in contrast to \figref{fig:p7Umap} where bin boundaries were used. The lower panel shows the resulting estimates for $\Rsun$ varying $\Vtot$ and allowing for $\Usun$ as free parameter. 
}\label{fig:p7rotangle}
\end{figure}

\subsection{Using the direction of motion}\label{sec:p7direct}

As a third strategy for fixing Galactic parameters I suggest the direction of motion. 
All that needs to be done is to compare the expected Galactic angle $\alpha$ to the angle that the Galactocentric mean motion in the subsamples has against the local azimuth. The upper panel in \figref{fig:p7rotangle} shows the angle 
\begin{equation}\label{eq:p7alphav}
\alpha_v = \arctan{\left(\frac{\langle U\rangle+\Usunzero}{\langle V \rangle + \Vtot}\right)}
\end{equation}
of the mean velocity for a binned sample of all stars with $\feh > -1.0$ (above that value I expect sufficient rotation) assuming a solar azimuthal velocity  $\Vtot = 248.5 \kms$, $\Usunzero = 13.0 \kms$ and plotting only bins with more than $10$ stars. The lines show positions of constant Galactic angle $\alpha$. For easier fits I note that the ratio $(\langle U\rangle+\Usunzero)/(\langle V \rangle + \Vtot)$ behaves as $y/(x+\Rsun)$ as long as we stay away from $|\alpha| \sim 0$ (avoiding the arctangent gives better convergence when starting far from the optimal parameters). It is obvious that this kind of statistic critically depends on the choice of $\Usun$ and the total azimuthal velocity of the Sun. 
Formally the agreement between the position angle and the velocity angle is optimised by fitting
\begin{equation}
\alpha'(\Rsun, \Usun) =\arctan{\left(\frac{y}{x + \Rsun}\right)} + \gamma(\Usun)
\end{equation}
\begin{equation}
\gamma(\Usun) = \arctan{\left(\frac{\langle U\rangle+\Usunzero}{\langle V \rangle + \Vtot}\right)} - \arctan{\left(\frac{\langle U\rangle+\Usun}{\langle V\rangle + \Vtot}\right)}
\end{equation}
to $\alpha_v$, where $\gamma(\Usun)$ does the correction from the first guess $\Usunzero$ to the parametrised solar radial velocity $\Usun$ and $x$ and $y$ are the coordinates of the local Cartesian frame in the radial (outwards) and azimuthal direction.

While I can fit $\Usun$ directly to these data, this is not possible for $V_{\rm tot}$ as the fit would then converge to the wrong global minimum: Let $V_{\rm tot}$ and $\Rsun$ go to infinity and the fit becomes perfect with all ratios/angles zeroed. In this light I vary $V_{\rm tot}$ externally and allow for $\Usun$ and $\Rsun$ as free fit parameters. All bins were weighted by the inverse number of stars they contain and I only accepted bins with more than $10$ objects. I note that velocity dispersions potentially exert a (predictable) distortion on $\alpha_v$ by its non-linearity, but found this effect to be very small on the current sample.

It is not surprising that the result for the solar radial velocity is $\Usun = (13.84 \pm 0.27) \kms$ in excellent agreement with the direct measurement from stellar $U$ velocities and with Section \ref{sec:p7systerr}. 

The result for $\Rsun$ is shown in the lower panel of \figref{fig:p7rotangle}, where I plot the new datapoints from angle fitting in addition to the previously discussed data. The formal error on $\Rsun$ at fixed $\Vtot$ is a small $0.12 \kpc$, but their relation largely resembles that of Sgr $A^{*}$. I would need a larger sample extension and/or significantly more stars to compete with Sgr $A^{*}$, which will be achieved with Gaia. For now, formal inclusion of the third constraint hence does not change the final results and just reduces the formal errors a bit. Nevertheless the potential of this estimator for larger samples is evident. The residuals of the fit will likely outline perturbations of the Galactic potential and substructure with data from Gaia, but in the present sample I could not detect any structure. 

Adding stars with halo metallicities to the test just drives up the error bars by a factor of $2$ because of the higher dispersion, while no rotating stars are added. $\Rsun$ stays the same while $\Usun$ shifts to $(14.10 \pm 0.41) \kms$. Restricting the disc sample to a distance closer than $2 \kpc$ increases $\Rsun$ and $\Usun$ by insignificant $0.06 \kpc$ and $0.06 \kms$. The value for $\Usun$ depends on the position of the Galactic centre. If our coordinate system was off-centre by $1$ degree in Galactic longitude, $\Usun$ could shift down by about $3 \kms$.

\section{Conclusions}\label{sec:p7conclude}

The most important outcomes of this work are three modelling-free and simple estimators for the Solar azimuthal velocity and hence the local circular speed $V_c$ and for the Solar Galactocentric radius $\Rsun$.

On this course I developed the idea that in a spatially extended sample the absolute rotation of stellar components can be measured from systematic streaming in the heliocentric radial direction. The stars on one side of the Galactic centre show an opposite heliocentric $U$ velocity to those on the other side. This value has a lower formal precision than the classically used azimuthal velocity, but can boost accuracy by its relative independence from assumptions about the velocity of the Sun. 

The rotation in any extended sample severely affects determinations of the Local Standard of Rest: $\Usun$ and $\Wsun$ are frequently determined via simple sample averages in each component. The rotation bias in $\Usun$ gets more important with increasing distance and impacts all presently available big surveys, since they are lopsided, i.e. asymmetric in Galactic longitude. For SEGUE dwarfs it amounts to $\sim 10 \kms$. Accounting for rotation the otherwise observed difference between disc and halo stars disappears and combining this classic method with the estimate from the mean direction of motion described below, I find $\Usun = (14.0 \pm 0.3)\kms$ with an additional systematic uncertainty of about $1.5 \kms$. The value is $\sim 3 \kms$ larger than from the Geneva-Copenhagen Survey, but still within the error margin. While the GCS is clearly affected by stellar streams, the presented value may be distorted by the problematic Sloan proper motions and possibly residual distance errors.

While there could of course be some interesting physics involved, a systematic difference of $\sim 4 \kms$ in the average $W$ motion between cones towards the Galactic North and South Poles points to a systematic error in the line-of-sight velocities by $\sim 2 \kms$. This is not implausible in light of the adhoc shift of $7.3 \kms$ applied in \citep[][]{SloanDR6,SloanDR8}. The correction reconciles $\Wsun$ to reasonable agreement with Hipparcos and the Geneva-Copenhagen Survey \citep[][]{Holmberg09,AB09} albeit at a lower value of about $6 \kms$.

Comparison of the absolute rotation measure $\theta$ based on heliocentric $U$ velocities to the mean azimuthal velocities $V_g$ in a sample delivers the solar azimuthal velocity $\Vtot$. This measurement is correlated with the assumed Galactocentric radius $\Rsun$. Combining this relation with another datum like the proper motion of Sgr $A^{*}$ one can determine both $\Rsun$ and $\Vtot$. For DR8 I obtain $\Rsun = 8.26^{+0.37}_{-0.33} \kpc$ and $\Vtot = 250^{+11}_{-10} \kms$. By dissecting the sample via metallicity into slow and fast rotating subgroups I can independently infer the Galactocentric radius from their comparison: A larger $\Rsun$ reduces $\alpha$ and hence nearly proportionally increases $\theta$. Thus fast rotators experience a larger absolute change in the rotation speed $\theta$, putting their value of $\Vtot$ at odds with that from the slow rotators when $\Rsun$ is wrong. Enforcing consistency provides $\Rsun = 8.29^{+0.63}_{-0.54} \kpc$, and in combination with the simple rotation measure and the proper motion of Sgr $A^{*}$ I get $\Rsun = (8.27 \pm 0.29) \kpc$ and $\Vtot = (250 \pm 9) \kms$ in excellent agreement with the values from \cite{McMillan11} or \cite{Gillessen09}. The circular speed $V_c = \Vtot - \Vsun$ using the LSR value of $\Vsun = (12.24 \pm 0.47 \pm 2) \kms$ from \cite{SBD} is then $V_c = (238 \pm 9) \kms$.

The third approach uses just the direction of motion in the Galaxy to estimate $\Rsun$. Assuming near axisymmetry again the direction of motion in a sample should always point in the direction of the azimuth. Fitting the Galactic angle throughout the plane to the angle of motion provides $\Rsun$. This measurement displays a strong dependence on the solar radial motion apart from the dependence on $\Vtot$. Fortunately $\Usun$ and $\Rsun$ are not degenerate and I obtain again a relatively large $\Usun = (13.84 \pm 0.27) \kms$, confirming the previous conventional analysis, but with an even smaller formal error. As the resulting relationship between $\Vtot$ and $\Rsun$ is only weakly inclined against the result from Sgr $A^{*}$, this approach will be of relevance for larger samples and here just provides some reassurance. 

The laid out strategies are extremely simple. They do not rely on any modelling with possible hidden assumptions, so that possible biases are readily understood. They rely on the sole assumption of near axisymmetry of the Galactic disc. They are designed for thick disc stars that reside high above the Galactic plane and are by their position and high random energy far less affected by perturbations of the Galactic potential, like the spiral pattern. Their modelling independence is supported by the fact that by construction none of the estimators depends on the detailed changes of sample composition with position in the Galaxy or on the radial dependence of $V_c$. Still some uncertainties must be pointed out: The methods are vulnerable to large-scale distortions from axisymmetry in the Galactic disc. The bar region should be avoided, and there may be a residual signal from spiral arms. I could not detect significant structures connected to this, so the consequences should be rather small, particularly as I stay out of the region dominated by the bar and as the sample is sufficiently extended not to cover just one side of a spiral arm. 

The currently available data add to the possible biases by their uncertain distances, radial velocities and especially astrometry. The Segue proper motions display a distinctive systematic pattern on quasar samples. While I use the quasar catalogues for corrections, the solution remains unsatisfactory: I could not quantify separately the physical reasons, i.e. chromatic aberrations, astrometric ``frame-dragging'' by stars mistaken for Galaxies or even focal plane distortions of the telescopes, and also there might be some undetected dependence on the stellar colours. Further some minor uncertainty in parameters is caused by possible biases in line-of-sight velocities.

I made extensive use of the distance corrections developed by \cite{SBA}. Besides that this project would have been futile without the accuracy achieved by the statistical corrections, the outcome is prone to all the weaknesses of that method. Especially streams and wrong assumptions about the velocity ellipsoid can induce systematic distance errors of order $5 \%$. The contamination in the sample will vary to some extent with Galactic coordinates, an effect that I did not model. This can lead to a bias, because the distribution of weights in the rotation estimators and the distance estimator is different. The accuracy could be far better could I use more stars and had I better control of the systematic errors by another dataset. Part of this will be resolved by use of additional samples like RAVE. Fortunately the estimators show very different behaviour on the different biases: The proper motion problem affects almost exclusively the first estimator for rotation with a correction to $\Vtot$ of $\sim 6 \kms$, while the difference between fast and slow rotators is nearly untouched. Vice versa, the difference estimator reacts strongly to distance changes, while the first estimator does not - a $10 \%$ change shifts it by $\sim 2 \kms$. Despite their benign distribution the systematic uncertainties on $\Vtot$ and $\Rsun$ may come close to the formal errors.

If the reader should take only one point from this note then let it be this: With the advent of the large surveys stellar samples are regaining their place as a primary source to obtain not only the Local Standard of Rest, but global Galactic parameters. Already a sample of $\sim 50000$ stars from SEGUE provides formal accuracies for the galactocentric radius and the solar azimuthal velocity that are competitive with any other known approach and without any need for modelling.

\section*{Acknowledgements}
It is a pleasure to acknowledge useful discussions with J. Binney, P. McMillan and M. Asplund on early drafts. I thank D. Weinberg for fruitful discussions and R. Dong for kind provision of the quasar samples and valuable discussions on Sloan proper motions. I acknowledge financial and material support from Max-Planck-Gesellschaft. Support for this work was partly provided by NASA through Hubble Fellowship grant $\#60031637$ awarded by the Space Telescope Science Institute, which is operated by the Association of Universities for Research in Astronomy, Inc., for NASA, under contract NAS 5-26555.

\label{lastpage}

\end{document}